\documentclass[journal,draftcls,onecolumn,12pt,twoside]{manuscript}
\usepackage{amsmath,amsfonts}
\usepackage{algorithmic}
\usepackage{algorithm}
\usepackage{array}
\usepackage[caption=false,font=normalsize,labelfont=sf,textfont=sf]{subfig}
\usepackage{textcomp}
\usepackage{stfloats}
\usepackage{url}
\usepackage{verbatim}
\usepackage{graphicx}
\usepackage{cite}
\usepackage{CJK}
\usepackage{indentfirst}
\usepackage{cases}
\usepackage{multirow}
\usepackage{rotating}
\usepackage{makecell}
\usepackage{paralist,bbding,pifont}
\usepackage{adjustbox}
\usepackage[percent]{overpic}

\begin{document}

\title{Unified Statistical Channel Modeling and performance analysis of Vertical Underwater Wireless Optical Communication Links considering Turbulence-Induced Fading}

\author{Dongling~Xu,~
        Xiang~Yi,~
        Yal\c{c}ın~Ata,~
        Xinyue~Tao,~
        Yuxuan~Li,~
        and~Peng~Yue
\thanks{This work was supported by National Key Research and Development Program of China (2022YFC2808101).}
\thanks{Dongling Xu, Xiang Yi, Xinyue Tao, Yuxuan Li and Peng Yue are with the State Key Laboratory of Integrated Service Networks, Xidian University, Xi'an 710071, China.Y. Ata is with Department of Electrical and Electronics Engineering, OSTIM Technical University, OSTIM, 06374 Yenimahalle, Ankara, Turkey. (email: donglingxu@stu.xidian.edu.cn;  yixiang@xidian.edu.cn; ylcnata@gmail.com; xyt@stu.xidian.edu.cn; lyuxuan@stu.xidian.edu.cn; pengy@xidian.edu.cn) (Corresponding Author: Peng Yue).}}

\markboth{Research Paper}%
{Research Paper}

\maketitle

\vspace{-5em}

\begin{abstract}
The reliability of a vertical underwater wireless optical communication (UWOC) network is seriously impacted by turbulence-induced fading due to fluctuations in the water temperature and salinity, which vary with depth. To better assess the vertical UWOC system performances, an accurate probability distribution function (PDF) model that can describe this fading is indispensable. In view of the limitations of theoretical and experimental studies, this paper is the first to establish a more accurate modeling scheme for wave optics simulation (WOS) by fully considering the constraints of sampling conditions on multi-phase screen parameters. On this basis, we complete the modeling of light propagation in a vertical oceanic turbulence channel and subsequently propose a unified statistical model named mixture Weibull-generalized Gamma (WGG) distribution model to characterize turbulence-induced fading in vertical links. Interestingly, the WGG model is shown to provide a perfect fit with the acquired data under all considered channel conditions. We further show that the application of the WGG model leads to closed-form and analytically tractable expressions for key UWOC system performance metrics such as the average bit-error rate (BER). The presented results give valuable insight into the practical aspects of development of UWOC networks.
\end{abstract}

\begin{IEEEkeywords}
Underwater wireless optical communication
(UWOC), vertical link, channel modeling, turbulence-induced fading, bit-error rate (BER).
\end{IEEEkeywords}

\IEEEpeerreviewmaketitle

\vspace{-1em}
\section{Introduction}

\IEEEPARstart{U}{nderwater} wireless optical communication (UWOC) systems have drawn a lot of attention due to their advantages of much higher data rate, greater security, and lower latency \cite{ref1}. Even though the transmission length is relatively short due to the light beam suffering from absorption, scattering, and turbulence-induced fading in sea water, UWOC is still a promising technology in many applications such as subsea observation networks to meet the growing demands for ocean exploration with high data rate transmission \cite{ref2}. 

Many existing investigations on the channel characteristics of the UWOC systems were mostly concerned with the inherent optical properties (IOPs) of seawater, i.e., absorption and scattering \cite{MCS1,MCS2,MCS3}. However, optical turbulence in ocean is also an indispensable factor that may cause performance deterioration of the UWOC systems \cite{ref4}. Optical turbulence, also known as the refractive-index fluctuations, will distort the wave front of the light beam and thus lead to the irradiance fading. Optical turbulence in ocean stems from the small-scale fluctuations of temperature as well as salinity. This makes it much stronger than that in atmosphere, which is mainly induced by the temperature variations \cite{ref3}. 

The reliability of an IM/DD UWOC system in oceanic turbulence can be evaluated from a mathematical model for the probability density function (PDF) of the irradiance fading. In this context, the accuracy and tractability of the PDF model is crucial in the estimation of the IM/DD UWOC system performance. For simplicity, the majority of the early work directly adopted PDF models proposed for atmospheric turbulence \cite{ref7,ref8,ref9,ref10,ref11}. They followed the same mathematical forms as that in atmosphere, but used the scintillation coefficient of oceanic turbulence instead. Specifically, by replacing the PDF parameters with the plane and spherical wave scintillation coefficients derived from Nikishovs spectrum \cite{ref40}, the Log-normal model was applied to investigate the average bit error rate (BER) of UWOC system in weak oceanic turbulence \cite{ref7}. Along this line of analysis, the K model and the Gamma-Gamma ($\Gamma \Gamma $) model described for moderate to strong turbulence \cite{ref4} were used to calculate the outage probability \cite{ref8,ref9}. And the Malaga \cite{ref10} as well as Exponentiated Weibull (EW) model \cite{ref11}, which can define a wider range of turbulence state, were later employed to evaluate the average channel capacity of UWOC system. Despite the contribution of these research efforts, extreme necessity have remained to explore the applicability of existing PDF models in the ocean. Furthermore, the complexity of wave equation and refractive-index spectrum becomes intractable to find the analytical solutions for the PDF parameters.   

To find a proper PDF model, a series of experiments have been conducted by scholars to emulate the oceanic turbulence. They used heating devices to create temperature gradients and injected sodium chloride into the tank to simulate salinity changes. By recording the fluctuation of the output signal, the PDF can be back-fitted \cite{ref12,ref13,ref14,ref15,ref16,ref17,ref18}. Oubei et al. utilized the generalized Gamma (GG) distribution to reflect the turbulence effects caused by temperature gradients \cite{ref12}. They then complemented the analysis of salinity change and used the Weibull distribution to characterize the fading \cite{ref13}. Nonetheless, this salinity change is only embodied in concentration rather than gradient. The impact of air bubbles-induced refractive-index fluctuations on the behavior of light propagation was further discussed in \cite{ref14}. Subsequently, the mixed Exponential-Gamma distribution \cite{ref15}, Exponential-Lognormal \cite{ref16}, Exponential-generalized Gamma (EGG) \cite{ref17} and Normal-Beta distributions \cite{ref18} were proposed to model light fading in UWOC channels. However, different from turbulence, which varies continuously with spatial position, air bubbles are dispersedly distributed in seawater, resulting in spatial discreteness in the refractive index fluctuations they induce \cite{ref19}. This makes it important to distinguish between the effects of turbulence and bubbles on the light beams. Additionally, it is challenging to reconstruct the actual state of the ocean in a lab environment owing to the complexity of the environment, both geographically and seasonally. 

A promising solution is to use Fourier optics theory-based wave optics simulation (WOS). This approach that enables low cost, short period, and rapid adjustment of parameters, has been developed for PDF modeling in turbulent media. The PDF models fitted based on WOS statistics are forcefully supported by the hydrodynamics and wave optics theories, making their accuracy and validity guaranteed. Relevant researches have been reported over the years \cite{ref20,ref21,ref22}. Zhang et al. and Shishter et al. selected the Log-normal distribution \cite{ref20} and Gamma distribution \cite{ref21} to fit the simulation data, respectively. Gao et al. applied the Weibull distribution to model the fading effect under various turbulent situations \cite{ref22}. While progress has been made in PDF modeling based on WOS, the simulation scenarios are still confined to atmospheric turbulence, other than that, the simulation process and the details of parameter design are not provided. 

The typical developments outlined above are all focused on horizontal channel PDF modeling, while there are numerous practical applications that necessitate the establishment of links between the deep ocean and the sea surface for data transfer. In this case, UWOC channels are generally vertical links. Hierarchical intensity cascading has been applied for PDF modeling of underwater vertical channel recently \cite{ref23,ref24,ref25,ref26,ref27,ref28,ref29,ref30}. The vertical link is divided into a number of closely connected and independent sub-paths, with the PDF of each sub-path being directly taken from the existing models, such as Log-normal \cite{ref23,ref24,ref25}, $\Gamma \Gamma $ \cite{ref26,ref27,ref28}, EW \cite{ref29} and mixed EGG models \cite{ref30}. The complete PDF can then be obtained by multiplying the fading distribution of each sub-layer. However, turbulence has a continuous effect on the propagation of the light field, and this effect cannot be separated, nor can it be considered as completely independent. Thus, additional study is required to develop a more appropriate vertical UWOC channel modeling.   

Recent research on the PDF of vertical UWOC channels leveraging WOS can be found in \cite{ref31}. To replicate the propagation of light beams over vertical underwater turbulent channels, Argo datasets from real oceans \cite{ref33} were used to generate a number of phase screens along the vertical link, and the log-normal distribution was assumed to examine the BER performance. Although a general simulation process and some necessary parameter values are given, the rationality of the parameter settings has not been verified by the theory of optical propagation, which makes it difficult to guarantee the accuracy of the WOS process and simulation results. Motivated by these, we establish the modeling of light propagation in a vertical oceanic turbulence channel and completed the performance analysis of the UWOC system. The major contributions of this work are as follows: 

\begin{compactitem}
\item To make up for the defects of the previous researches, the WOS parameters are designed for the vertical communication scenario on basis of Fourier optics theory \cite{ref32}. This makes the simulation conform to the basic principles of optics and hydrodynamics, and thus ensures the correctness of the simulation results. 
\item In view of the correctness of the simulation, we establish the modeling of light propagation in a vertical oceanic turbulence channel and perform a range of simulation realizations. 
\item The correctness of simulation results guarantees the accuracy of statistical PDFs. On this foundation, we propose a unified statistical model named WGG distribution model, which can efficiently describe turbulence-induced fading under vast majority of channel conditions for vertical ocean, providing analytical tractability as well. 
\item Based on the WGG model, we evaluate the effects of average signal-to-noise ratio (SNR), optical transmitter depth, and link depth in different ocean areas on the average BER of the vertical UWOC system. 
\end{compactitem}

This study is the first to perform statistics on the light intensity fluctuations of vertical turbulent channels and to produce a distribution model with both theoretical rationality and experimental applicability, which is of great scientific significance for the performance analysis of vertical UWOC links. The remainder of the paper is organized as follows. Section II models the vertical underwater turbulent optical channel. Section III introduces the mixture WGG model and present the expectation maximization (EM) algorithm in detail. We describe the stratification conditions of seawater temperature and salinity with depth in Section IV. Section V and Section VI are devoted to the analysis of the light propagation characteristics and the error rate performance in the vertical oceanic turbulence links, respectively. Finally, Section VII concludes the paper. 

\section{Modeling of Vertical Underwater Turbulent Optical Channel}
\subsection{Multi-step WOS for optical signal propagation in vertical oceanic turbulent optical channel}
\begin{figure}[H]
\vspace{-2em}
\setlength{\abovecaptionskip}{0.cm}
\centering
\begin{overpic}[width=2.9in]{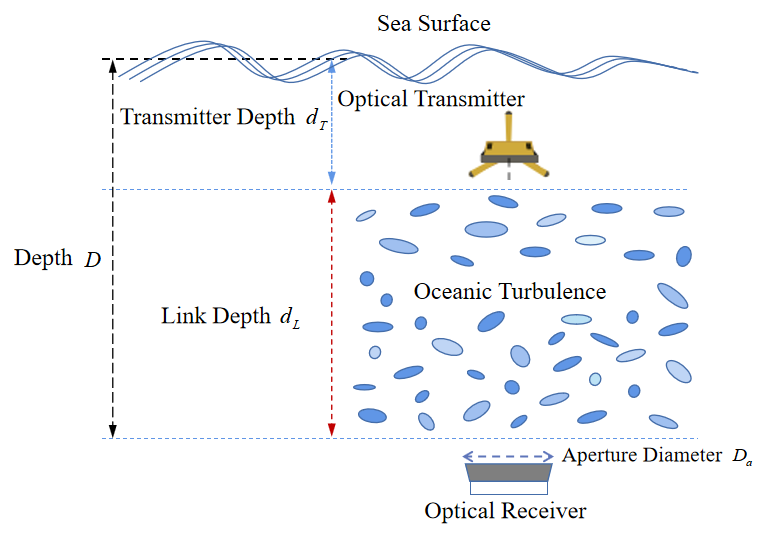}
\put(90,60){\small\textbf{(a)}}
\end{overpic}
\begin{overpic}[width=3.4in]{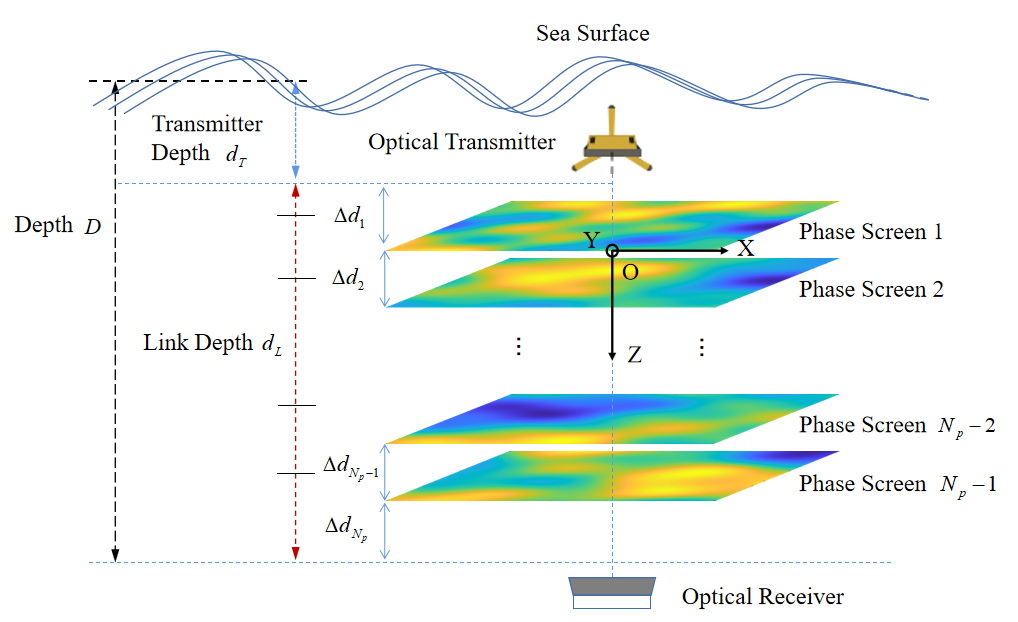}
\put(90,50){\small\textbf{(b)}}
\end{overpic}
\caption{(a) Vertical UWOC link under consideration. (b) Schematic of multi-phase screen method to simulate the transmission process of optical signals in vertical underwater turbulent channel.}
\label{fig1}
\vspace{-2em}
\end{figure}
As illustrated in Fig. \ref{fig1}(a), we consider an underwater vertical link with transceivers at different depth positions. The laser transmitter is assumed to be placed at a depth ${d_T}$ below sea surface and initially send a Gaussian beam, whose optical field can be expressed as \cite{ref4} 
\begin{equation}
\label{ex1}
U\left( {x,y,z = 0} \right) = \exp \left[ { - \left( {{x^2} + {y^2}} \right)/w_0^2} \right],
\end{equation}
where ${w_0}$ denotes the effective beam radius. Then the optical signal it emits passes through the vertically distributed turbulent medium and eventually reaches the optical receiver with an aperture ${D_a}$. ${d_L}$ represents the link depth between transceivers. In this way, the receiver is at the position of $D = {d_T} + {d_L}$ below the sea surface. 

The WOS uses the multi-step approach where the continuous turbulent path is subdivided into several layers and a phase screen is located at the center of each layer to account for the effect of turbulence within the segment volume. Different from horizontal oceanic channel, depth is a key factor in affecting the turbulence intensity of the vertical UWOC links. Because of this, all phase screens are placed oriented perpendicular to the direction of gravity for the vertical underwater turbulent optical channel, as shown in Fig. \ref{fig1}(b). The multi-step beam propagation method is to alternate steps of partial vacuum propagation with interaction between the light and the turbulence media to accomplish simulating propagation through oceanic turbulence. For this purpose, the propagation path is divided into ${N_p}$ equal intervals by ${N_p} - 1$ phase screens, and the optical field in the final plane can be written as \cite{ref32} 
\begin{equation}
\label{ex2}
\begin{array}{l}
U\left( {{x_{{N_p}}},{y_{{N_p}}}} \right) =\\ Q\left[ {\frac{{{m_{{N_p} - 1}} - 1}}{{{m_{{N_p} - 1}}\Delta {d_{{N_p} - 1}}}},\left( {{x_{{N_p}}},{y_{{N_p}}}} \right)} \right] \left\{ {Q\left[ {\frac{{1 - {m_1}}}{{\Delta {d_1}}},\left( {{x_1},{y_1}} \right)} \right]{\cal T}\left[ {{z_1},{z_2}} \right]U\left( {{x_1},{y_1}} \right)} \right\}\times \prod\limits_{p = 1}^{{N_p} - 1}\\
{\left\{ {\frac{{{\cal T}\left[ {{z_p},{z_{p + 1}}} \right]}}{{{m_p}}}{{\cal F}^{ - 1}}\left[ {\left( {{\kappa _{xp}},{\kappa _{yp}}} \right),\frac{{\left( {{x_{p + 1}},{y_{p + 1}}} \right)}}{{{m_p}}}} \right]} \right.} \left. {{Q_2}\left[ { - \frac{{\Delta {d_p}}}{{{m_p}}},\left( {{\kappa _{xp}},{\kappa _{yp}}} \right)} \right]{\cal F}\left[ {\left( {{x_{p + 1}},{y_{p + 1}}} \right),\left( {{\kappa _{xp}},{\kappa _{yp}}} \right)} \right]} \right\},
\end{array}
\end{equation}
where ${\cal F}$ and ${{\cal F}^{ - 1}}$ stand for Fourier transform (FT) and inverse Fourier transform (IFT), respectively. ${\cal T}\left[ {{z_p},{z_{p + 1}}} \right] = \exp \left[ { - i{\varphi _{p}}\left( {x,y} \right)} \right]$ is an operator denoting the accumulation of phase perturbation induced by the oceanic turbulence with ${\varphi _p}\left( {x,y} \right)$ being distribution function of the ${p^{th}}$ random phase screen. $Q\left[  \cdot  \right]$ and ${Q_2}\left[  \cdot  \right]$ are the operator notations of phase factor defined by Nazarathy and Shamir \cite{ref35}
\begin{equation}
\label{ex3}
Q\left[ {\frac{{1 - {m_p}}}{{\Delta {d_p}}},\left( {{x_1},{y_1}} \right)} \right] = {e^{ - \frac{{ik}}{2}\left( {\frac{{1 - {m_p}}}{{\Delta {d_p}}}} \right)\left( {x_1^2 + y_1^2} \right)}},\quad {Q_2}\left[ { - \frac{{\Delta {d_p}}}{{{m_p}}},\left( {{\kappa _{xp}},{\kappa _{yp}}} \right)} \right] = {e^{ - \frac{{ik}}{2}\left( {\frac{{1 - {m_p}}}{{\Delta {d_p}}}} \right)\left( {x_1^2 + y_1^2} \right)}},
\end{equation}
where $k = 2\pi /\lambda $ denotes wave number with $\lambda $ being light wavelength. $\left( {{\kappa _{xp}},{\kappa _{yp}}} \right)$ is the spatial wave number vector of ${p^{th}}$ phase screen. ${m_p}$ in \eqref{ex2} defined as ${m_p} = {\delta _p}/{\delta _{p - 1}}$ denotes the scaling factor from ${\left( {p - 1} \right)^{th}}$ plane to ${p^{th}}$ plane, where ${\delta _{p - 1}}$ and ${\delta _p}$ are the grid spacing in the corresponding phase screens. $\Delta {d_p}$ represents interval length between adjacent planes. It is worth noting that the set phase screen parameters need to meet certain criteria, which are discussed in the following section. 
\subsection{Constraints on the random phase screen parameters}
In the WOS, it is critical to appropriately pick the grid spacing and number of grid points to ensure an accurate simulation. This determines the number of phase screens set on the propagation path and the maximum interval length between adjacent phase screens. For two adjacent phase screens (the first two phase screens as an example), we recall the Nyquist criterion to place a constraint on the grid spacing $\delta $ such that $\delta  \le 1/\left( {2{f_{\max }}} \right)$, where ${f_{\max }}$ is the maximum spatial frequency of interest \cite{ref32}. The key to achieving an accurate result is to sample the quadratic phase factor inside the FT at a high enough rate to satisfy the Nyquist criterion. Local spatial frequency ${{\bf{f}}_{loc}}$ is the local rate of change of a waveform given by ${{\bf{f}}_{loc}} = \nabla \phi /\left( {2\pi } \right)$, where $\nabla $ denotes the gradient operator. $\phi $ is the optical phase. We need to find the maximum local spatial frequency of the quadratic phase factor inside the integral and sample at least twice this rate to ensure that all of the present spatial frequencies are not aliased \cite{ref32}.  

In \eqref{ex3}, the phase of the first phase factor is $\phi  = \frac{k}{2}\left( {\frac{{1 - m}}{{\Delta d}}} \right)\left( {x_1^2 + y_1^2} \right)$. Since the quadratic phase has the same variations in the both Cartesian directions, we just analyze the ${x_1}$ direction, which yields the local spatial frequency ${f_l}$ as 
\begin{equation}
\label{ex11}
{f_l} = \frac{1}{{2\pi }}\frac{\partial }{{\partial {x_1}}}\phi  = \frac{1}{\lambda }\left( {\frac{{1 - {\delta _2}/{\delta _1}}}{{\Delta d}}} \right){x_1}.
\end{equation}
The maximum spatial frequency occurs at ${x_1} = {D_1}/2$, in which ${D_1}$ denotes the maximum spatial extent of the source field. Applying the Nyquist sampling gives $\frac{1}{\lambda }\left( {\frac{{1 - {\delta _2}/{\delta _1}}}{{\Delta d}}} \right)\frac{{{D_1}}}{2} \le \frac{1}{{2{\delta _1}}}$. After some algebra, we obtain 
\begin{equation}
\label{ex13}
{\delta _1} - \frac{{\lambda \Delta d}}{{{D_1}}} \le {\delta _2} \le {\delta _1} + \frac{{\lambda \Delta d}}{{{D_1}}}.
\end{equation}
The phase of the second quadratic phase factor is $\phi  = {\pi ^2}\frac{{2\Delta d}}{{mk}}{\left| {{{\bf{f}}_1}} \right|^2}$, where $\left| {{{\bf{f}}_1}} \right| = \sqrt {\kappa _{x1}^2 + \kappa _{y1}^2} $. The local spatial frequency ${f_l}^\prime $ is derived as 
\begin{equation}
\label{ex15}
{f'_l} = \frac{1}{{2\pi }}\frac{\partial }{{\partial {f_1}}}\phi  = \frac{{{\delta _1}\lambda \Delta d}}{{{\delta _2}}}{f_1}.
\end{equation}
This is a maximum at the edge of the spatial-frequency grid where ${f_1} = 1/2{\delta _1}$. Applying Nyquist sampling criterion gives $\frac{{\lambda \Delta d}}{{2{\delta _2}}} \le \frac{1}{{2{\delta _f}}}$, where ${\delta _f}$ denoting the grid spacing in the spatial frequency domain of the phase screen, has the form ${\delta _f} = N{\delta _1}/2$ with $N$ representing the number of grid points. Eventually, we can obtain the constraints of $\Delta d$ given by 
\begin{equation}
\label{ex17}
\Delta d \le N{\delta _1}{\delta _2}/\lambda .
\end{equation}
Thus, the number of phase screens ${N_p}$ set in the propagation path is necessary to satisfy 
\begin{equation}
\label{ex18}
{N_p} \ge {d_L}/\Delta {d_{\max }},
\end{equation}
where $\Delta {d_{\max }}$ denotes the maximum length of the interval between adjacent phase screens. 
\begin{figure}[H]
\vspace{-1em}
\setlength{\abovecaptionskip}{0.cm}
\centering
\includegraphics[width=5in]{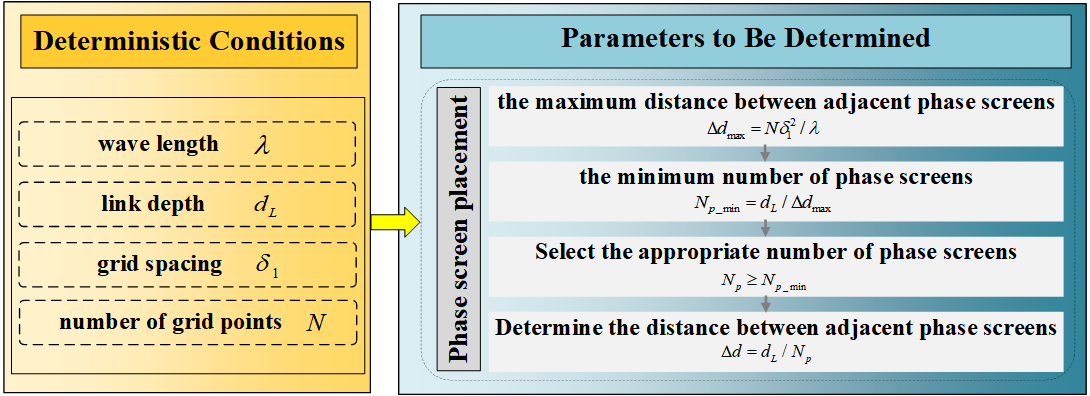}
\caption{Design for phase screen placement.}
\label{fig4}
\vspace{-2em}
\end{figure}

According to the above theoretical analysis, we present the design of the phase screen placement form shown in Fig. \ref{fig4}. For each WOS implementation, determining the depth position of the phase screen is a key step to ensure the accuracy of the simulation. Concretely, given the deterministic propagation conditions, such as optical wavelength, grid spacing and the number of grid points, the maximum distance between adjacent phase screens can be solved for based on \eqref{ex17}, and combined with the simulated link depth, the minimum number of phase screens can be calculated according to \eqref{ex18}. After that, we choose the appropriate number of phase screens subject to the inequality constraint. Considering that the number of grid points and the grid spacing are the same for each plane set in the simulation, the phase screens are set to be placed at equal intervals. At this point, the depth position of the phase screens is determined. 
\subsection{Phase screen generation based on oceanic turbulence spectrum model}
The power spectrum inversion method is the more commonly used method for turbulence phase screen generation. Given the low-frequency compensation, the distribution function of the multiple oceanic turbulence phase screen is given by \cite{ref38} 
\begin{equation}
\label{ex19}
\begin{array}{l}
{\kern 1pt} {\varphi _p}(u{\delta _x},v{\delta _y}) = {\left[ {{{\left( {\frac{{2\pi }}{N}} \right)}^2}\frac{1}{{{\delta _x}{\delta _y}}}} \right]^{\frac{1}{2}}}\left\{ {\sum\limits_{{u^\prime }} {\sum\limits_{{v^\prime }} h } \left( {{u^\prime },{v^\prime }} \right)\sqrt {{F_{\varphi ,p}}\left( {{u^\prime },{v^\prime }} \right)} \exp \left[ {i2\pi \left( {\frac{{{u^\prime }u}}{N} + \frac{{{v^\prime }v}}{N}} \right)} \right]} \right.\\
\left. {+ \sum\limits_{\ell  = 1}^N {\sum\limits_{{u^\prime } =  - 1}^1 {\sum\limits_{{v^\prime } =  - 1}^1 h } } \left( {{u^\prime },{v^\prime }} \right)\sqrt {{F_{\varphi ,p}}\left( {{u^\prime },{v^\prime }} \right)} \exp \left[ {i2\pi \left( {\frac{{{u^\prime }u}}{{{3^l}N}} + \frac{{{v^\prime }v}}{{{3^l}N}}} \right)} \right]} \right\},
\end{array}
\end{equation}
where $u,v,{u^\prime },{v^\prime }$ are integers. ${\delta _x}$ and ${\delta _y}$ are the x- and y-directed sampling interval. For the sake of simplicity, we set ${\delta _x} = {\delta _y} = {\delta _1}$. $h\left( {{u^\prime },{v^\prime }} \right)$ represents complex Gaussian statistic. ${F_{\varphi ,p}}\left(  \cdot  \right)$ is phase power spectrum density of the oceanic turbulence, which is related to spatial refractive index spectrum ${\Phi _{n,p}}\left(  \cdot  \right)$ by ${\kern 1pt} {F_{\varphi ,p}}\left( {{\kappa _x},{\kappa _y}} \right) = 2\pi {k^2}\Delta {d_p}{\Phi _{n,p}}(\kappa )$ \cite{ref4}.

In this paper, the spatial refractive index spectrum reported in \cite{ref39}, which varies with environmental parameters, has been applied for correlation analysis and can be expressed as 
\begin{equation}
\label{ex23}
{\kern 1pt} {\Phi _{n,p}}(\kappa ) = A_p^2{\Phi _{T,p}}(\kappa ) + B_p^2{\Phi _{S,p}}(\kappa ) + 2{A_p}{B_p}{\Phi _{TS,p}}(\kappa ).
\end{equation}
Here ${A_p}$ and ${B_p}$ are the linear coefficients varying with the water’s average temperature $\left\langle T \right\rangle $ (Kelvin) and salinity $\left\langle S \right\rangle $ (ppt) for the ${p^{th}}$ layer. ${\Phi _{T,p}}(\kappa )$, ${\Phi _{S,p}}(\kappa )$ and ${\Phi _{TS,p}}(\kappa )$ are the temperature spectrum, the salinity spectrum, and the co-spectrum, respectively. These three spectra can be expressed uniformly as follows \cite{ref39} 
\begin{equation}
\label{ex26}
\begin{array}{l}
{\kern 1pt} {\Phi _{q,p}}(\kappa ) = \frac{{{\beta _0}{\chi _{q,p}}{\varepsilon ^{ - 1/3}}}}{{4\pi }}{\kappa ^{ - 11/3}}\exp \left[ { - 174.90{{\left( {\kappa {\eta _p}} \right)}^2}c_{q,p}^{0.96}} \right]\\
 \times \left[ {1 + 21.61{{\left( {\kappa {\eta _p}} \right)}^{0.61}}c_{q,p}^{0.02} - 18.18{{\left( {\kappa {\eta _p}} \right)}^{0.55}}c_{q,p}^{0.04}} \right],\quad q \in \left\{ {T,S,TS} \right\},
\end{array}
\end{equation}
where ${\beta _0}$ is the Oboukhov-Corrsin constant. ${c_{q,p}}$ is the non-dimensional parameter associated with $\left\langle T \right\rangle $ and $\left\langle S \right\rangle $, which can be calculated by using the (23) in \cite{ref39}. $\varepsilon $ represents the rate of dissipation of kinetic energy per unit mass of fluid. ${\eta _p}$ denotes the Kolmogorov microscale. The details regarding variation of ${\eta _p}$ with $\left\langle T \right\rangle $ and $\left\langle S \right\rangle $ are given in the Appendix A of \cite{ref39}. The ensemble-averaged variance dissipation rates ${\chi _{q,p}}\left( {q \in \left\{ {T,S,TS} \right\}} \right)$ are expressed as \cite{ref40,ref41} 
\begin{equation}
\label{ex27}
{\chi _{T,p}} = {K_T}\left( {\frac{{\partial T}}{{\partial z}}} \right)_p^2,\quad{\chi _{S,p}} = \frac{{\alpha _{c,p}^2{\chi _{T,p}}d{r_p}}}{{\left( {\omega _p^2\beta _{c,p}^2} \right)}},\quad{\chi _{TS,p}} = \frac{{{\alpha _{c,p}}{\chi _{T,p}}\left( {1 + d{r_p}} \right)}}{{2{\omega _p}{\beta _{c,p}}}},
\end{equation}
where ${K_T}$ indicates the dispersion coefficient of turbulence in the vertical direction \cite{ref31}. $d{r_p}$ is defined as the eddy diffusivity ratio of salinity to temperature of the ${p^{th}}$ layer, expressed as (9) of \cite{ref42}, and ${\omega _p}$ denotes the ratio of temperature and salinity contributions to the refractive index spectrum \cite{ref39}, which can be expressed as ${\omega _p} = \frac{{{\alpha _{c,p}}}}{{{\beta _{c,p}}}}\left| {{H_p}} \right|$, where ${H_p}$ is the temperature-salinity gradient ratio, which can be expressed as ${\left( {\partial T/\partial z} \right)_p}/{\left( {\partial S/\partial z} \right)_p}$. ${\alpha _{c,p}}$ and ${\beta _{c,p}}$ represent the thermal expansion coefficient and the saline contraction coefficient that can be calculated by TEOs-10 toolbox \cite{ref34,ref54}. 
\begin{figure}[!htp]
\vspace{-1.5em}
\setlength{\abovecaptionskip}{-0.3cm}
\centering
\includegraphics[width=5in]{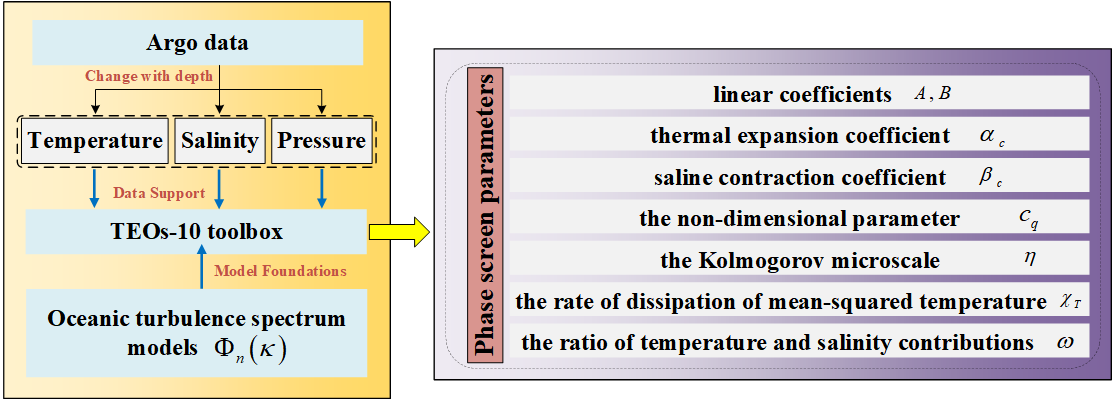}
\caption{Design for phase screen parameters.}
\label{fig5}
\vspace{-2em}
\end{figure}

We summarize the design process for the phase screen parameters generation in Fig. \ref{fig5}. In the underwater vertical link, the turbulent channel parameters are commonly functions of the oceanic inherent thermodynamic properties such as temperature, salinity, and pressure. The fluctuations in these oceanic thermodynamic are pertinent to depth, which means that the different depths will change the underwater turbulent channel state. Therefore, according to the data support provided by Argo datasets from real oceans, combined with selected turbulence power spectrum, TEOs-10 toolbox can be used to obtain the channel parameters, thereby generating the distribution function of the random phase screen for the corresponding depth. 

\section{Modeling Underwater Turbulence-Induced Intensity Fluctuation with The Mixture WGG Model}
\subsection{Statistics of the unified model}
In this paper, we proposed using the mixture WGG distribution, which is a weighted sum of the Weibull and GG distributions to model the turbulence-induced fading in vertical UWOC link. The mixture WGG distribution can be expressed as 
\begin{equation}
\label{ex43}
{f_I}\left( I \right) = \varpi f\left( {I;\left[ {\beta ,\tilde \eta } \right]} \right) + \left( {1 - \varpi } \right)g\left( {I;\left[ {\tilde a,\tilde d,\tilde p} \right]} \right),
\end{equation}
with
\begin{equation}
\label{ex44}
f\left( {I;\left[ {\beta ,\tilde \eta } \right]} \right) = \frac{\beta }{{\tilde \eta }}{\left( {\frac{I}{{\tilde \eta }}} \right)^{\beta  - 1}}\exp \left[ { - {{\left( {\frac{I}{{\tilde \eta }}} \right)}^\beta }} \right],g\left( {I;\left[ {\tilde a,\tilde d,\tilde p} \right]} \right) = \frac{{\tilde p{I^{\tilde d - 1}}}}{{{{\tilde a}^{\tilde d}}\Gamma \left( {\tilde d/\tilde p} \right)}}\exp \left[ { - {{\left( {I/\tilde a} \right)}^{\tilde p}}} \right].
\end{equation}
$f$ and $g$ are respectively the Weibull and GG distributions. $\varpi $ is the mixture weight or mixture coefficient of the distributions, satisfying $0 < \varpi  < 1$. $\beta $ and $\tilde \eta$ are the parameter associated with the Weibull distribution. $\tilde a$, $\tilde d$ and $\tilde p$ are the parameters of the GG distribution. 

\subsection{Maximum Likelihood (ML) parameter estimation of the WGG model}
Throughout this paper, we apply the EM algorithm to find the mixture WGG model parameters that realize the best fit with the simulation data. The EM algorithm, generally developed for ML estimation of models involving missing data, has also been applied to estimate the parameters of mixture models \cite{ref17}. For a specific channel condition, we set 2000 simulation realizations and collect $1024 \times 1024$ intensity fluctuations data samples for each simulation realization. It is possible to associate each observed intensity realization ${I_o}$ with a hidden unobserved binary random variable ${\tilde z_o}$ taking 1 with probability $\varpi $ when the data point is drawn from the Weibull distribution and 0 with probability $1 - \varpi $ if drawn from the GG distribution. The EM-algorithm seeks to determine the ML estimates of the parameters of the mixture model in \eqref{ex43} by alternating the following two steps.  

E-step: The E-step consists in computing the expected values of the hidden variables $\left\{ {{{\tilde z}_o}} \right\}$ given the incomplete data set $\left\{ {{I_o}} \right\}_{o = 1}^{Num}$. Using the Bayes’ rule, these quantities are given by 
\begin{equation}
\label{ex47}
\begin{array}{l}
{\gamma _o} \buildrel \Delta \over = {\rm E}\left[ {{{\tilde z}_o} = 1\left| {\left\{ {{I_o}} \right\}_{o = 1}^{Num}} \right.} \right] = \frac{{\varpi f\left( {{I_o};\left[ {\beta ,\tilde \eta } \right]} \right)}}{{\varpi f\left( {{I_o};\left[ {\beta ,\tilde \eta } \right]} \right) + \left( {1 - \varpi } \right)g\left( {{I_o};\left[ {\tilde a,\tilde d,\tilde p} \right]} \right)}}.
\end{array}
\end{equation}

M-step: The M-step consists in selecting the parameters of the mixture model that maximize the following functional which coincides with the expected value of the log likelihood function of the complete data set $\left\{ {\left( {{I_o},{{\tilde z}_o}} \right)} \right\}_{o = 1}^{Num}$ with respect to the conditional distribution $\left( {{{\tilde z}_1},...,{{\tilde z}_{Num}}} \right)$ given ${I_1},...,{I_{Num}}$, 
\begin{equation}
\label{ex48}
\begin{array}{l}
\ell \left( {{I_o};\left[ {\beta ,\tilde \eta } \right],\left[ {\tilde a,\tilde d,\tilde p} \right]} \right) = \sum\limits_{o = 1}^{Num} {\left[ {{\gamma _o}\ln \left( {f\left( {{I_o};\left[ {\beta ,\tilde \eta } \right]} \right)} \right) + {\gamma _o}\ln \varpi } \right.} \\
\left. { + \left( {1 - {\gamma _o}} \right)\ln \left( {g\left( {{I_o};\left[ {\tilde a,\tilde d,\tilde p} \right]} \right)} \right) + \left( {1 - {\gamma _o}} \right)\ln \left( {1 - \varpi } \right)} \right].
\end{array}
\end{equation}

For Weibull distribution, taking the derivatives of functional $\ell $ with respect to $\tilde \eta $ and $\beta $ leads to the following set of equations 
\begin{equation}
\label{ex49}
\tilde \eta  = \sum\limits_{o = 1}^{Num} {{\gamma _o}} /\left( {\sum\limits_{o = 1}^{Num} {{\gamma _o}I_o^\beta } } \right),
\end{equation}
\begin{equation}
\label{ex50}
\sum\limits_{o = 1}^{Num} {{\gamma _o}\left( {\frac{1}{\beta } + \ln {I_o} - \tilde \eta I_o^\beta \ln {I_o}} \right)}  = 0.
\end{equation}
To find $\beta $, it suffices to replace $\tilde \eta $ with its expression in \eqref{ex49} into \eqref{ex50}. Then, $\tilde \eta $ can be retrieved using again \eqref{ex49}. 

As for GG distribution, it is handier to work with $\vartheta  = {\tilde a^{\tilde p}},{\rm{ }}\tilde q = \tilde d/\tilde p$. Taking the derivatives of functional $\ell $ with respect to $\vartheta $, $\tilde p$ and $\tilde q$ results in 
\begin{equation}
\label{ex51}
\vartheta  = \frac{{\sum\limits_{o = 1}^{Num} {\left( {1 - {\gamma _o}} \right)I_o^{\tilde p}} }}{{\sum\limits_{o = 1}^{Num} {\left( {1 - {\gamma _o}} \right)\tilde q} }}, \quad\tilde q = \frac{{\sum\limits_{o = 1}^{Num} {\frac{{\left( {1 - {\gamma _o}} \right)}}{{\tilde p}}} }}{{\frac{{\sum\limits_{o = 1}^{Num} {\left( {1 - {\gamma _o}} \right)I_o^{\tilde p}\ln {I_o}} \sum\limits_{o = 1}^{Num} {\left( {1 - {\gamma _o}} \right)} }}{{\left( {\sum\limits_{o = 1}^{Num} {\left( {1 - {\gamma _o}} \right)I_o^{\tilde p}} } \right)}} - \sum\limits_{o = 1}^{Num} {\left( {1 - {\gamma _o}} \right)\ln {I_o}} }},
\end{equation}
\begin{equation}
\label{ex53}
\sum\limits_{o = 1}^{Num} {\left( {1 - {\gamma _o}} \right)\left( {\tilde p\ln {I_o} - \ln \vartheta  - \frac{{\Gamma '\left( {\tilde q} \right)}}{{\Gamma \left( {\tilde q} \right)}}} \right)}  = 0.
\end{equation}
Similar to the treatment of the Weibull distribution, once $\tilde p$ is found, $\vartheta $ and $\tilde q$ can be logically obtained by use of \eqref{ex51}. Eventually, the weight $\varpi $ satisfies $\varpi  = \sum\limits_{o = 1}^{Num} {{\gamma _o}} /Num$. 

\subsection{Goodness of Fit Tests}
The validity of the introduced unified model may be verified statistically by conducting goodness of fit (GoF) tests that describe the degree of accordance between the PDF model and the simulation data. More specifically, the mean square error (MSE) test and the R-square (${R^2}$) test that have been widely employed in evaluating the goodness of fit of a variety of fading distributions to channel collection data are adopted in this paper. 

\subsubsection{${R^2}$ Test}
The coefficient of determination ${R^2}$, which is used to quantify the goodness of fit, can be expressed as \cite{ref50} 
\begin{equation}
\label{ex55}
{R^2} = 1 - \sum\nolimits_{\nu  = 1}^M {{{\left( {{f_{m,\nu }} - {f_{p,\nu }}} \right)}^2}} /\sum\nolimits_{\nu  = 1}^M {{{\left( {{f_{m,\nu }} - \bar f} \right)}^2}} ,
\end{equation}
in which $M$ is the number of bins of the acquired data histogram, ${f_{m,\nu }}$ and ${f_{p,\nu }}$ are respectively the collected and predicted probability values for a given received intensity level corresponding to the ${\nu ^{th}}$ bin, and $\bar f = \sum\nolimits_{\nu  = 1}^M {{f_{m,\nu }}/M} $. The ${R^2}$ measure ranges from 0 to 1, and as ${R^2} \to 1$, the distribution better fits the acquired data. 

\subsubsection{MSE Test}
The MSE is a simple and efficient measure, defined as 
\begin{equation}
\label{ex56}
{\mathop{\rm MSE}\nolimits}  = \left[ {\sum\nolimits_{o = 1}^{Num} {{{\left( {{F_e}\left( {{I_o}} \right) - F\left( {{I_o}} \right)} \right)}^2}} } \right]/Num,
\end{equation}
where ${F_e}\left(  \cdot  \right)$ represents the empirical distribution function and $F\left(  \cdot  \right)$ stands for the theoretical cumulative distribution function computed with parameters estimated from the collected data. It is worth mentioning that ${\mathop{\rm MSE}\nolimits}  \to 0$ indicates a better fit to the acquired simulation data. 

\section{Variation Trends of Seawater Temperature and Salinity with Depth}
\begin{figure}[H]
\vspace{-2em}
\setlength{\abovecaptionskip}{0.cm}
\centering
\begin{overpic}[width=2.55in]{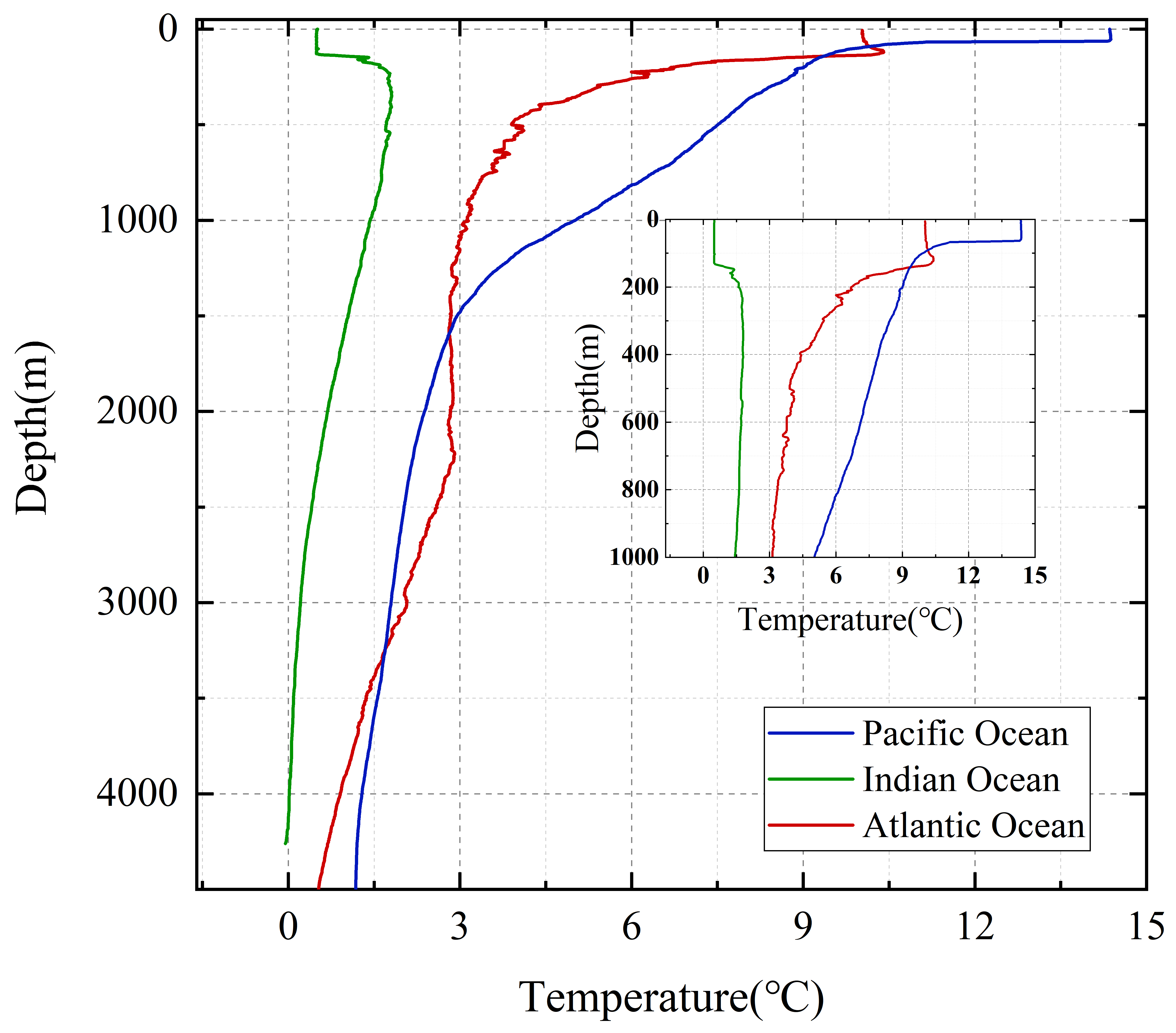}
\put(85,80){\small\textbf{(a)}}
\end{overpic}
\begin{overpic}[width=2.55in]{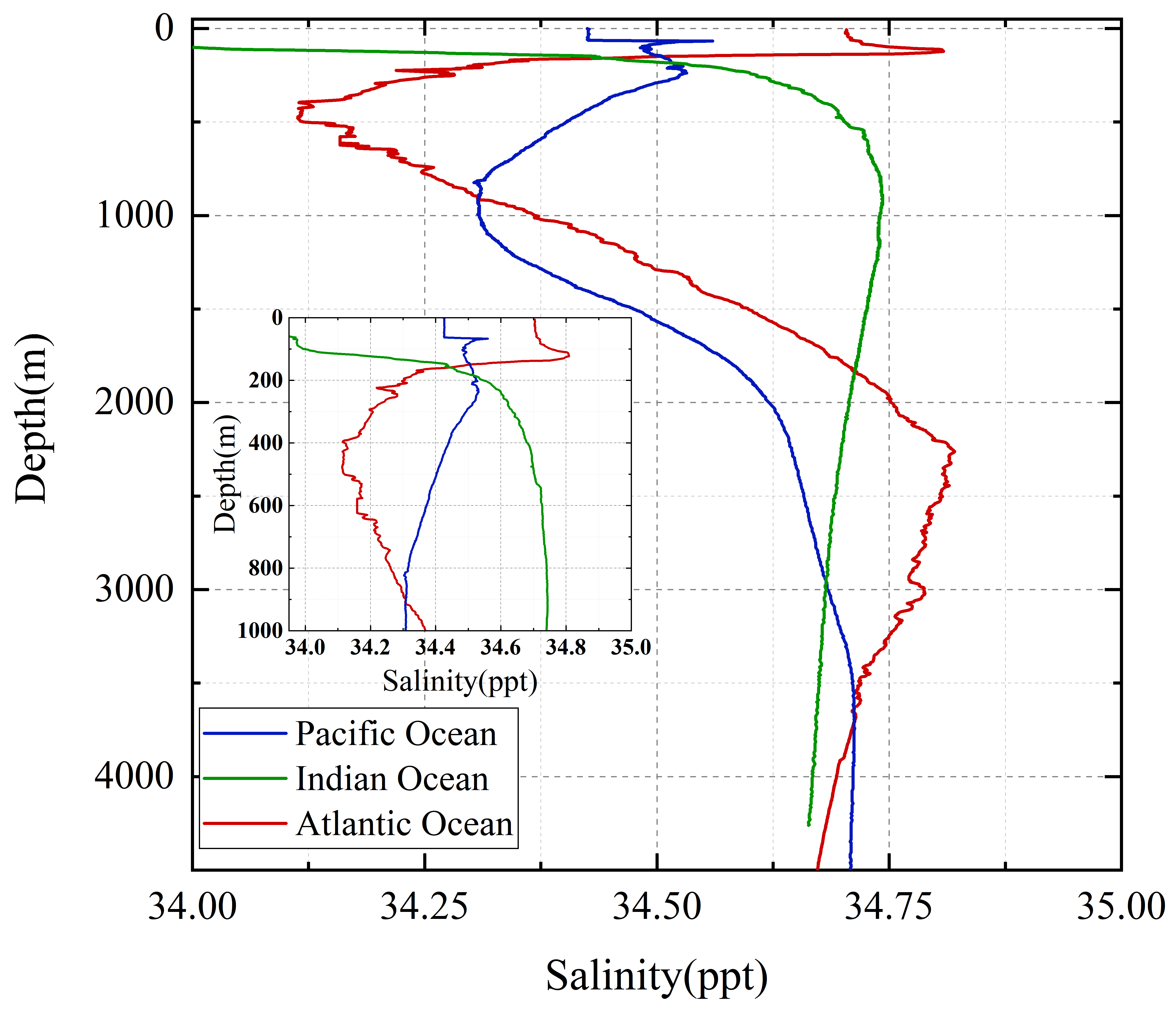}
\put(85,80){\small\textbf{(b)}}
\end{overpic}
\caption{Oceanic profiles. (a) Temperature. (b) Salinity.}
\label{fig6}
\vspace{-2em}
\end{figure}

\begin{table}[!h]
\vspace{-1.5em}
\begin{center}
\caption{Geographic location and date of Argo data acquisition employed in the simulation.}
\label{tab1}
\resizebox{!}{!}{
\renewcommand\arraystretch{0.7}
\begin{tabular}{ c  c  c  c  c }
\hline
Ocean Areas & ID of Nodes & Latitude & Longitude & Date \\
\hline
Pacific Ocean & 5906361 & -43.5428 & -150.0083 & 2023/04/16 \\
Indian Ocean & 7900679 & -60.1513 & 105.7122 & 2022/12/31 \\ 
Atlantic Ocean & 3901565 & -44.9764 & -51.8703 & 2022/08/20 \\
\hline 
\end{tabular}}
\end{center}
\vspace{-4.5em} 
\end{table}
The simulations employ the measured data of Argo buoy nodes in the Atlantic, Pacific and Indian Ocean, and the locations of these nodes are listed in Table \ref{tab1} \cite{ref33}. We plot the variation curves of temperature and salinity with depth, as shown in Fig. \ref{fig6}. It is obvious that both profiles show depth-dependency regardless of the ocean areas. Additionally, the depths of the mixed layers and thermocline are different in each ocean area. For instance, the measured data at node 7900679 and 3901565 show that the depth of the mixed layer in the Indian and Atlantic Ocean can reach more than 100 m, whereas that for node 5906361 in the Pacific Ocean is only a few tens of meters. It can be also noted that the range of change in temperature with depth is much larger than the range of change in salinity. 

\section{Analysis of The Propagation Characteristics of Optical Waves in A Vertical Oceanic Turbulence Channel}
Based on the model of the vertical channel for optical communication established in section II, the propagation characteristics simulation is performed. Unless otherwise stated, all the necessary simulation parameters of interest are listed in Table \ref{tab2}. After calculation, it is verified that 10 turbulence phase screens we set satisfies the constraint on the random phase screen parameters in section II(B). Moreover, 2000 realizations are used to collect each result data point. 
\begin{table}[!h]
\vspace{-2em}
\begin{center}
\caption{Simulation parameter setting.}
\label{tab2}
\resizebox{!}{!}{
\renewcommand\arraystretch{0.7}
\begin{tabular}{ c  c || c  c }
\hline
Parameters & Value & Parameters & Value\\
\hline
Grid spacing ${\delta _1}$ & 0.25 mm & Effective beam radius ${w_0}$ & 3 cm \\
Number of grid points $N$ & 1024 & Aperture diameter ${D_a}$ & 10 cm \\ 
Light wavelength $\lambda $ & 532 nm & Dispersion coefficient of turbulence ${K_T}$ & ${10^{ - 5}}$ ${{\rm{m}}^2}/{\rm{s}}$ \\
Communication link depth ${d_L}$ & 70 m & Turbulent energy dissipation rate $\varepsilon $ & ${10^{ - 5}}$ ${{\rm{m}}^2}/{\rm{s^3}}$ \\
Transmitter depth ${d_T}$ & 20 m & Number of phase screens ${N_p}$ & 10 \\
\hline 
\end{tabular}}
\end{center}
\vspace{-5em} 
\end{table}

\subsection{Light intensity evolutionary properties}
In this section, we presented the numerical results for the effects of vertical oceanic turbulence on the intensity evolutionary properties of light propagating in different ocean areas. 
\begin{figure}[!htbp]
\vspace{-1.5em}
\setlength{\abovecaptionskip}{-0.cm}
\centering
\begin{overpic}[width=2.75in]{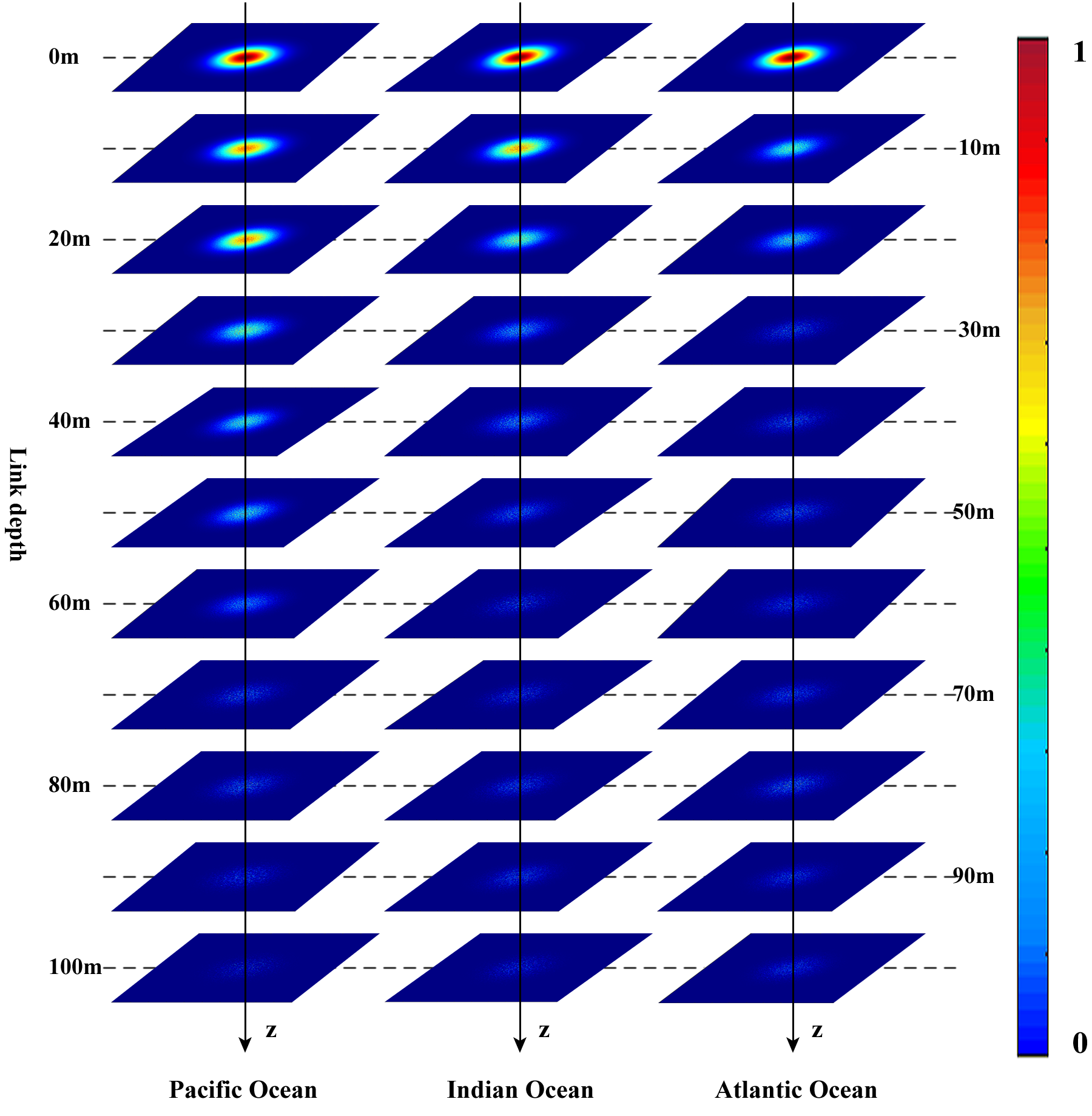}
\put(100,90){\small\textbf{(a)}}
\end{overpic}
\hspace{6mm}
\begin{overpic}[width=2.75in]{fig7b}
\put(100,90){\small\textbf{(b)}}
\end{overpic}
\caption{The normalized intensity images of Gaussian beam propagating in different ocean areas with different link depth. (a) 10m. (b) 100m.}
\label{fig7}
\vspace{-2.5em}
\end{figure}

According to the information for the three sea areas is given by Table \ref{tab1}, Fig. \ref{fig7} illustrates the normalized intensity evolution of Gaussian beam at different depths along the propagation in vertical turbulence channel. The transmitter is set at underwater with a depth of 20 m. Two link depths are considered: 10 m and 100 m, as shown in Fig. \ref{fig7}(a) and Fig. \ref{fig7}(b), respectively. Fig. \ref{fig7}(a) shows that as the propagation distance increases, the main flap energy of the Gaussian beam keeps spreading from inward to outward. This evolution phenomenon of the Gaussian beam is more obvious when beam propagates in the Atlantic Ocean. From Fig. \ref{fig7}(b), we can see that when the propagation depth exceeds a certain range, the beam spot is gradually broken, making it difficult to count its size. For the sake of description, we define this depth range where the spot size can be counted as the effective propagation depth. It can be seen that the effective propagation depths of Gaussian beams in the Pacific, Indian and Atlantic Oceans are 50 m, 20 m and 20 m, respectively. This difference can be analyzed by reviewing Fig. \ref{fig6}, specifically, the location of the transmitter is in the mixed layer, where the surface seawater moves violently due to tides, waves and currents, making the turbulence intensity significantly stronger and the impact on the effective propagation depth of the beam more severe. It can be seen that the mixed layer depth in the Indian and Atlantic Oceans is obviously greater than that in the Pacific Ocean, and accordingly, the beam is affected by turbulence in a larger depth range, which may result in a smaller effective propagation depth. 

\subsection{Normalized intensity statistical distribution}
Throughout this subsection, we take into account a variety of scenarios that include different transmitter depths, receiver apertures, propagation depths and ocean areas to evaluate the validity of various statistical distributions in predicting turbulence-induced fading in vertical UWOC channels. Meanwhile, we show how the unified WGG model provides an excellent agreement with the acquired simulation data under all channel conditions by compared with the Log-normal, Gamma, K, Weibull, EW, $\Gamma \Gamma $, GG as well as the EGG distributions, whose expressions and parameter notation can be found in \cite{ref14,ref17}. 

\begin{figure}[!htbp]
\vspace{-0.5em}
\setlength{\abovecaptionskip}{-0.3cm}
\centering
\includegraphics[width=5.5in]{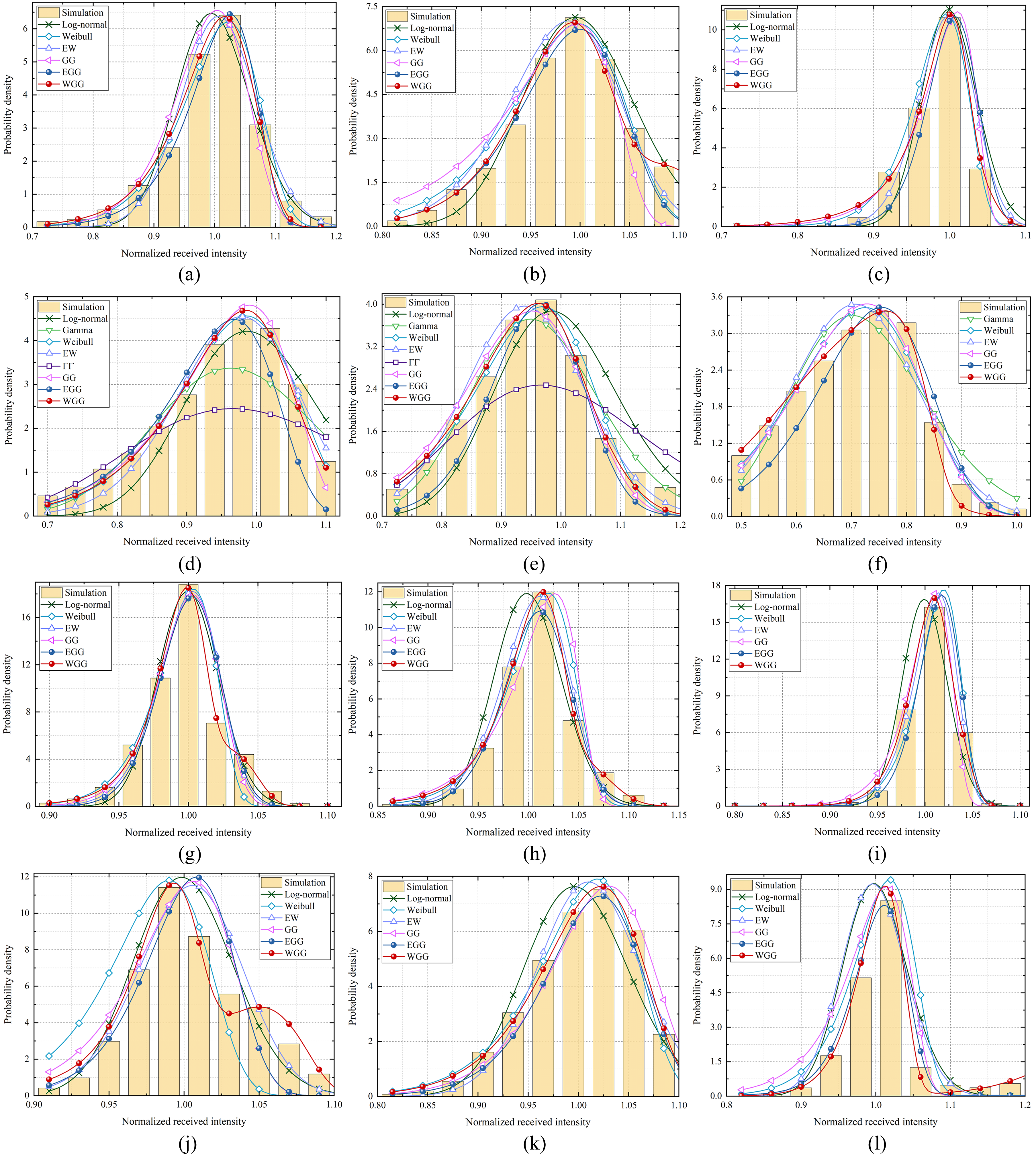}
\caption{Accordance of different statistical distributions, mentioned in section III, with the PDF histograms of the acquired simulation data through various UWOC channel conditions in Pacific Ocean.}
\label{fig8}
\vspace{-3.5em}
\end{figure}

\begin{table}
\begin{center}
\caption{GoF and the constant parameters of different PDFs for the various channel conditions considered in Pacific Ocean.}
\label{tab3}
\resizebox{15cm}{5cm}{
\renewcommand\arraystretch{1.5}
\begin{tabular}{|| c || c || c || c || c || c || c || c || c || c || c ||}
\hline

\multicolumn{2}{|| c ||}{\multirow{3}{*}{\textbf{Channel Condition}}} & \textbf{Log-normal} & \textbf{Gamma} & \textbf{K} & \textbf{Weibull} & \textbf{EW} & \textbf{$\Gamma \Gamma $} & \textbf{GG} & \textbf{EGG} & \textbf{WGG} \\
\cline{3-11} 

\multicolumn{2}{|| c ||}{\multirow{2}{*}{}} & ${R^2}$, MSE & ${R^2}$, MSE & ${R^2}$, MSE & ${R^2}$, MSE & ${R^2}$, MSE & ${R^2}$, MSE & ${R^2}$, MSE & ${R^2}$, MSE & ${R^2}$, MSE \\

\multicolumn{2}{|| c ||}{} & (${\mu _X},\sigma _X^2$) & ($\tilde k,\theta $) & ($\alpha $) & ($\beta ,\tilde \eta $) & ($\alpha ,\beta ,\tilde \eta $) & ($\alpha ,\beta $) & ($\tilde a,\tilde d,\tilde p$) & ($\tilde \omega ,\tilde \lambda ,\tilde a,\tilde d,\tilde p$) & ($\varpi ,\beta ,\tilde \eta ,\tilde a,\tilde d,\tilde p$) \\ 
\hline

\multirow{5}{*}{\makecell{${d_L} = 10$ m, \\ ${D_a} = 6$ cm}} & ${d_T} = 50$ m & \makecell{0.9152, 0.0076 \\  ($-9.588\times {10^{ - 4}}$, \\ $9.588\times {10^{ - 4}}$)} & --,-- & --,-- & \makecell{0.9680, \\ $5.7707\times {10^{ - 4}}$ \\ (17.292, 1.026)} & \makecell{0.9410, 0.0061 \\ (6.589, 7.621, 0.902)}  & --,-- & \makecell{0.9553, 0.0027 \\ (0.9911, 20.012, \\ 15.531)} & \makecell{0.9622, $8.0171\times {10^{ - 4}}$ \\ (0.1002, 6.4962, 1.0396, \\ 18.0091, 23.0208)} & \makecell{0.9890, $1.0487\times {10^{ - 4}}$ \\ (0.7531, 19.581, 1.029, \\ 1.014, 12.0169, 23.8298)} \\
\cline{2-11} 

& ${d_T} = 500$ m & \makecell{0.9419, $4.5570\times {10^{ - 4}}$ \\  ($-7.87\times {10^{ - 4}}$, \\ $7.87\times {10^{ - 4}}$)} & --,-- & --,-- & \makecell{0.8831, 0.0049 \\ (18.801, 1.001)} & \makecell{0.9194, $5.3267\times {10^{ - 4}}$ \\ (2.436, 12.197, 0.955)} & --,-- & \makecell{0.7516, 0.0100 \\ (1.029, 12.854, \\ 34.672)} & \makecell{0.8971, 0.0048 \\ (0.1035, 21.3739, 0.9894, \\ 22.5387, 17.8796)} & \makecell{0.9837, $4.1244\times {10^{ - 4}}$ \\ (0.7212, 22.932, 0.991, \\ 1.109, 16.243, 34.693)} \\
\cline{2-11} 

& ${d_T} = 1000$ m & \makecell{0.9486, 0.0030 \\  ($-3.272\times {10^{ - 4}}$, \\ $3.272\times {10^{ - 4}}$)} & --,-- & --,-- & \makecell{0.9759, 0.0013 \\ (29.291, 0.9969)} & \makecell{0.9738, 0.0022 \\ (3.732, 15.672, 0.959)} & --,-- & \makecell{0.9366, 0.0034 \\ (1.029, 21.079, \\ 52.721)} & \makecell{0.9216, 0.0036 \\ (0.1102, 24.0638, 0.9539, \\ 52.2896, 18.6553)} & \makecell{0.9782, $6.4424\times {10^{ - 4}}$ \\ (0.4215, 41.982, 1.002, \\ 1.0242, 16.245, 28.132)} \\
\hline 

\multirow{5}{*}{\makecell{${d_L} = 100$ m, \\ ${D_a} = 6$ cm}} & ${d_T} = 50$ m & \makecell{0.7300, 0.0191 \\  ($-2.282\times {10^{ - 4}}$, \\ $2.282\times {10^{ - 4}}$)} & \makecell{0.7700, 0.0083 \\ (67.421, 0.0145)} & --,-- & \makecell{0.9737, 0.0016 \\ (12.262, 0.9923)} & \makecell{0.9193, 0.0043 \\ (2.729, 7.473, 0.921)} & \makecell{0.3419, 0.0243 \\ (72.462, 72.672)} & \makecell{0.9762, $7.6209\times {10^{ - 4}}$ \\ (1.028, 10.271, \\ 18.429)} & \makecell{0.8186, 0.0053 \\ (0.1083, 41.3621, 0.9982, \\ 11.1179, 17.1483)} & \makecell{0.9767, $5.4205\times {10^{ - 4}}$ \\ (0.8042, 11.346, 0.9932, \\ 1.0034, 15.375, 21.832)} \\
\cline{2-11} 

& ${d_T} = 500$ m & \makecell{0.6730, 0.0290 \\  ($-2.711\times {10^{ - 4}}$, \\ $2.711\times {10^{ - 4}}$)} & \makecell{0.9593, 0.0039 \\ (79.421, 0.0121)} & --,-- & \makecell{0.9616, 0.0019 \\ (10.475, 0.979)} & \makecell{0.9618, 0.0012 \\ (4.238, 5.227, 0.831)} & \makecell{0.5375, 0.0300 \\ (75.622, 72.868)} & \makecell{0.9397, 0.0055 \\ (0.949, 10.632, \\ 9.374)} & \makecell{0.7758, 0.0206 \\ (0.2182, 16.8241, 0.8324, \\ 22.4192, 7.2736)} & \makecell{0.9655, 0.0010 \\ (0.6923, 9.482, 0.981, \\ 0.981, 11.487, 15.932)} \\
\cline{2-11} 

& ${d_T} = 1000$ m & --,-- & \makecell{0.9164, \\ $4.3165\times {10^{ - 4}}$ \\ (34.721, 0.0208)} & --,-- & \makecell{0.9440, \\ $1.9604\times {10^{ - 4}}$ \\ (6.832, 0.7424)} & \makecell{0.9458, $1.3769\times {10^{ - 4}}$ \\ (2.241, 4.547, 0.661)} & --,-- & \makecell{0.9434, $2.1930\times {10^{ - 4}}$ \\ (0.752, 6.767, \\ 7.277)} & \makecell{0.8513, 0.0085 \\ (0.1131, 12.9252, 0.7688, \\ 7.9328, 8.0476)} & \makecell{0.9657, $1.1120\times {10^{ - 4}}$ \\ (0.6311, 5.793, 0.682, \\ 0.822, 9.738, 19.073)} \\
\hline 

\multirow{5}{*}{\makecell{${d_L} = 10$ m, \\ ${D_a} = 10$ cm}} & ${d_T} = 50$ m & \makecell{0.9273, $5.2260\times {10^{ - 4}}$ \\  ($-1.201\times {10^{ - 4}}$, \\ $1.201\times {10^{ - 4}}$)} & --,-- & --,-- & \makecell{0.9089, 0.0011 \\ (50.231, 1.004)} & \makecell{0.9038, 0.0023 \\ (2.583, 31.291, 0.988)} & --,-- & \makecell{0.9122, $6.3616\times {10^{ - 4}}$ \\ (0.981, 68.0119, \\ 32.432)} & \makecell{0.9058, 0.0017 \\ (0.0241, 16.7578, 0.9041, \\ 118.7234, 18.0829)} & \makecell{0.9943, $2.778\times {10^{ - 4}}$ \\ (0.6351, 64.763, 0.998, \\ 1.039, 28.218, 56.187)} \\
\cline{2-11} 

& ${d_T} = 500$ m & \makecell{0.8786, 0.0052 \\  ($-2.81\times {10^{ - 4}}$, \\ $2.81\times {10^{ - 4}}$)} & --,-- & --,-- & \makecell{0.9348, 0.0033 \\ (33.109, 1.021)} & \makecell{0.9644, 0.0028 \\ (2.601, 20.291, 0.9879)} & --,-- & \makecell{0.8766, 0.0072 \\ (1.0419, 25.0128, \\ 51.429)} & \makecell{0.9793, 0.0017 \\ (0.0991, 17.3899, 0.9145, \\ 66.3655, 14.7647)} & \makecell{0.9905, $4.2209\times {10^{ - 4}}$ \\ (0.6023, 41.801, 1.0129, \\ 1.0419, 20.022, 24.023)} \\
\cline{2-11} 

& ${d_T} = 1000$ m & \makecell{0.9372, 0.0011 \\  ($-1.394\times {10^{ - 4}}$, \\ $1.394\times {10^{ - 4}}$)} & --,-- & --,-- & \makecell{0.8629, 0.0051 \\ (49.012, 1.0201)} & \makecell{0.9554, $9.2215\times {10^{ - 4}}$ \\ (1.4203, 40.041, 1.0099)} & --,-- & \makecell{0.9884, $7.0165\times {10^{ - 4}}$ \\ (1.0165, 42.594, \\ 56.0734)} & \makecell{0.8715, 0.0040 \\ (0.0478, 7.3183, 0.9886, \\ 74.9064, 30.4812)} & \makecell{0.9889, $2.6663\times {10^{ - 4}}$ \\ (0.374, 50.514, 1.028, \\ 1.0071, 52.0168, 55.8221)} \\
\hline 

\multirow{5}{*}{\makecell{${d_L} = 100$ m, \\ ${D_a} = 10$ cm}} & ${d_T} = 50$ m & \makecell{0.9025, 0.0015 \\  ($-2.78\times {10^{ - 4}}$, \\ $2.78\times {10^{ - 4}}$)} & --,-- & --,-- & \makecell{0.6647, 0.0131 \\ (31.785, 0.991)} & \makecell{0.8198, 0.0017 \\ (4.621, 15.672, 0.961)} & --,-- &  \makecell{0.7826, 0.0023 \\ (1.008, 31.902, \\ 32.532)} & \makecell{0.7466, 0.0030 \\ (0.1024, 52.6123, 0.9903, \\ 45.7878, 27.4486)} & \makecell{0.9445, $4.5002\times {10^{ - 4}}$ \\ (0.4875, 50.603, 0.992, \\ 1.072, 18.782, 48.832)} \\
\cline{2-11} 

& ${d_T} = 500$ m & \makecell{0.8468, 0.0025 \\  ($-6.87\times {10^{ - 4}}$, \\ $6.87\times {10^{ - 4}}$)} & --,-- & --,-- & \makecell{0.9621, \\ $4.8994\times {10^{ - 4}}$ \\ (21.912, 1.0212)} & \makecell{0.9576, 0.0020 \\ (7.8204, 8.81, 0.912)} & --,-- &  \makecell{0.9616, 0.0017 \\ (1.012, 25.011, \\ 17.523)} & \makecell{0.9715, $2.2621\times {10^{ - 4}}$ \\ (0.1125, 4.5251, 0.9866, \\ 29.7653, 16.2055)} & \makecell{0.9859, $1.6643\times {10^{ - 4}}$ \\ (0.5011, 23.574, 1.041, \\ 1.0068, 22.0171, 20.8313)} \\
\cline{2-11} 

& ${d_T} = 1000$ m & \makecell{0.7929, 0.0283 \\  ($-4.654\times {10^{ - 4}}$, \\ $4.654\times {10^{ - 4}}$)} & --,-- & --,-- & \makecell{0.8141, 0.0260 \\ (26.102, 1.0201)} & \makecell{0.7846, 0.0306 \\ (8.589, 9.991, 0.9071)} & --,-- &  \makecell{0.8630, 0.0251 \\ (1.032, 20.014, \\ 35.423)} & \makecell{0.9684, $4.2892\times {10^{ - 4}}$ \\ (0.2622, 5.6031, 0.9949, \\ 37.1925, 24.2629)} & \makecell{0.9759, $2.8789\times {10^{ - 4}}$ \\ (0.7023, 35.821, 1.013, \\ 1.307, 20.992, 30.822)} \\
\hline 
\end{tabular}}
\end{center}
\vspace{-5em}
\end{table}

Fig. \ref{fig8} shows the histograms of the acquired simulation data along with the fitness of the WGG probability distribution function under different vertical turbulence channel conditions in Pacific Ocean. Figs. \ref{fig8}(a)-(f) set the receiver aperture diameter to 6 cm, while Figs. \ref{fig8}(g)-(l) set it to 10 cm. We can clearly observe that the light intensity distribution with a small aperture is more dispersed than that with a large aperture. In fact, increasing the receiver aperture to larger than that of the spatial scale of the intensity fluctuations, will significantly reduce the signal fluctuations through averaging the fluctuations over the aperture which is well known as aperture averaging effect \cite{ref51}. For the transmitter location, we simulated three transmitter depths with ${d_T} = 50$ m, 500 m, 1000 m. It can be seen that with all other conditions remaining the same, the histogram corresponding to the channel environment with ${d_T} = 1000$ m tends to perform more concentrated. A likely explanation is that compared with the mixed layer and thermocline, the change of temperature-salinity gradient in the deep layer is low, so the light intensity suffers less perturbation. Additionally, the corresponding histograms of the acquired simulation data depicted in Fig. \ref{fig8} demonstrate that increasing the link depth significantly exacerbates the degree of light intensity fluctuation. This is due to the fact that the turbulence intensity is positively related to the propagation path length in a certain range. 

To better visualize the fit of the PDFs to the data, detailed simulation results for the GoF values of each of the discussed statistical distributions in section III and their corresponding constant coefficients are listed in Table \ref{tab3}. It shows that Log-normal, Weibull, EW, GG and EGG distributions fit the acquired data quite well when the link depth in the vertical channel is shallow (Figs. \ref{fig8}(a)-(c) and Figs. \ref{fig8}(g)-(i)). As the link depth increases, these distributions do not seem to follow the stretching shape of the graph very well. However, the proposed WGG model perfectly matches the acquired data under all channel conditions. It is clearly illustrated that the MSE values associated with the WGG model are the smallest, and the corresponding ${R^2}$ measures have the highest values. These results indicate that the unified PDF provides a better fit to the simulation data. 

\begin{figure}[!htbp]
\vspace{0em}
\setlength{\abovecaptionskip}{-0.3cm}
\centering
\includegraphics[width=5.5in]{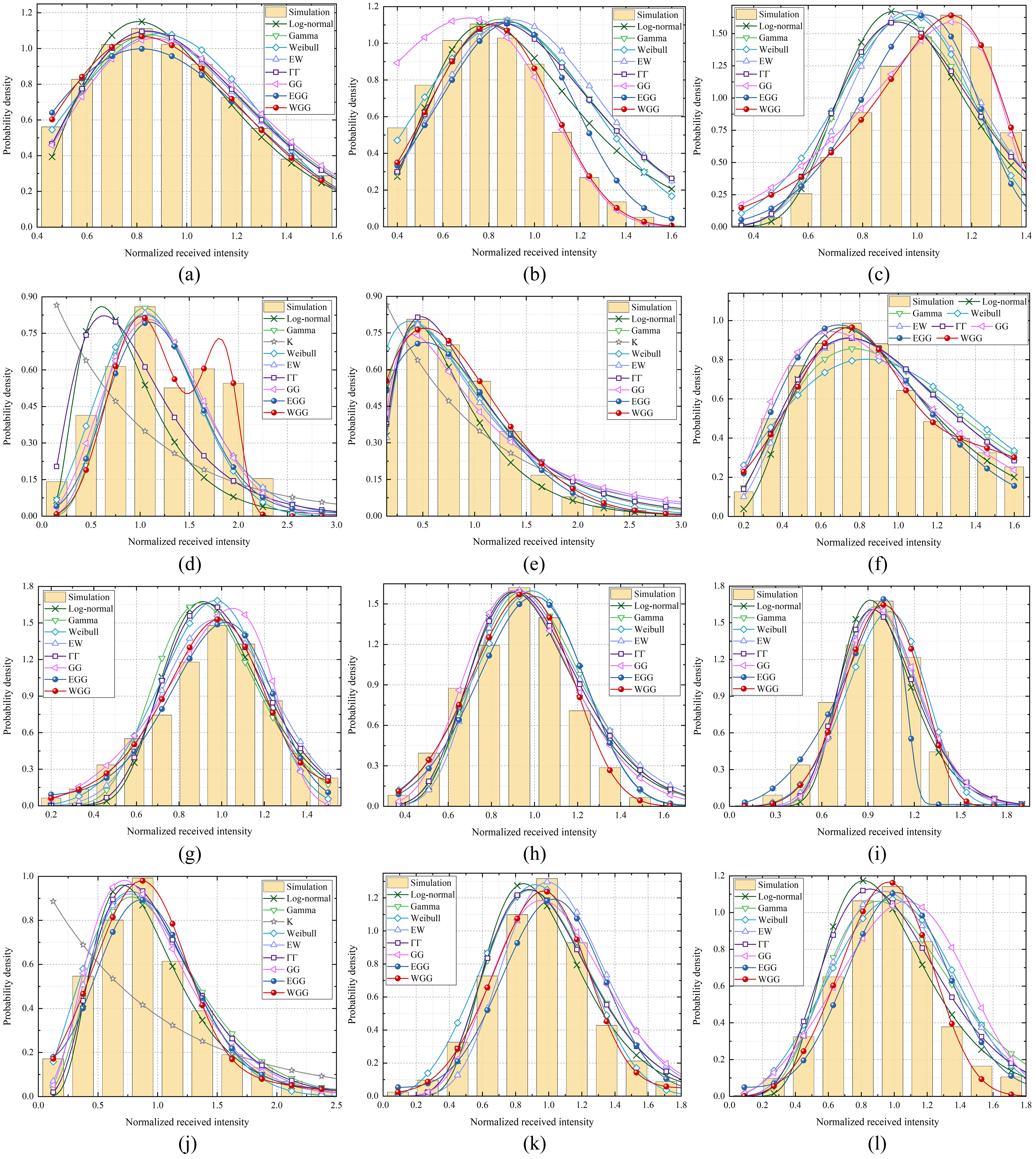}
\caption{Accordance of different statistical distributions, mentioned in section III, with the PDF histograms of the acquired simulation data through various UWOC channel conditions in Indian Ocean.}
\label{fig9}
\vspace{-3.5em}
\end{figure}

\begin{table}
\begin{center}
\caption{GoF and the constant parameters of different PDFs for the various channel conditions considered in Pacific Ocean.}
\label{tab4}
\resizebox{15cm}{5cm}{
\renewcommand\arraystretch{1.5}
\begin{tabular}{|| c || c || c || c || c || c || c || c || c || c || c ||}
\hline

\multicolumn{2}{|| c ||}{\multirow{3}{*}{\textbf{Channel Condition}}} & \textbf{Log-normal} & \textbf{Gamma} & \textbf{K} & \textbf{Weibull} & \textbf{EW} & \textbf{$\Gamma \Gamma $} & \textbf{GG} & \textbf{EGG} & \textbf{WGG} \\
\cline{3-11} 

\multicolumn{2}{|| c ||}{} & ${R^2}$, MSE & ${R^2}$, MSE & ${R^2}$, MSE & ${R^2}$, MSE & ${R^2}$, MSE & ${R^2}$, MSE & ${R^2}$, MSE & ${R^2}$, MSE & ${R^2}$, MSE \\

\multicolumn{2}{|| c ||}{} & (${\mu _X},\sigma _X^2$) & ($\tilde k,\theta $) & ($\alpha $) & ($\beta ,\tilde \eta $) & ($\alpha ,\beta ,\tilde \eta $) & ($\alpha ,\beta $) & ($\tilde a,\tilde d,\tilde p$) & ($\tilde \omega ,\tilde \lambda ,\tilde a,\tilde d,\tilde p$) & ($\varpi ,\beta ,\tilde \eta ,\tilde a,\tilde d,\tilde p$) \\ 
\hline

\multirow{5}{*}{\makecell{${d_L} = 10$ m, \\ ${D_a} = 6$ cm}} & ${d_T} = 50$ m & \makecell{0.8974, \\ $7.7264\times {10^{ - 4}}$ \\  (-0.0401, 0.0401)} & \makecell{0.9093, \\ $6.3577\times {10^{ - 4}}$ \\ (6.409, 0.157)} & --,-- & \makecell{0.8561, 0.0039 \\ (2.861, 1.041)} & \makecell{0.9311, \\ $3.9424\times {10^{ - 4}}$ \\ (4.299, 1.384, 0.594)} & \makecell{0.9194, \\ $3.9626\times {10^{ - 4}}$ \\ (165.732, 6.802)}  & \makecell{0.8652, 0.0011 \\ (0.3561, 5.106, \\ 1.293)} & \makecell{0.9348, $2.3319\times {10^{ - 4}}$ \\ (0.0142, 7.2194, \\ 0.8820, 2.8204, 2.1441)} & \makecell{0.9594, $1.7511\times {10^{ - 4}}$ \\ (0.3436, 2.128, 1.016, \\ 0.534, 4.495, 1.662)} \\
\cline{2-11} 

& ${d_T} = 500$ m & \makecell{0.8436, 0.0039 \\  (-0.045, 0.045)} & \makecell{0.6552, 0.0070 \\  (6.993, 0.1421)} & --,-- & \makecell{0.7309, 0.0061 \\ (2.842, 1.0132)} & \makecell{0.5341, 0.0080 \\ (1.8836, 2.168, \\ 0.889)} & \makecell{0.6634, 0.0064 \\ (159.572, 7.036)} & \makecell{0.8679, 0.0025 \\ (1.084, 1.626, \\ 4.934)} & \makecell{0.8890, 0.0017 \\ (0.2285, 0.2285, \\ 1.0391, 3.2373, 4.1497)} & \makecell{0.9776, 0.0011 \\ (0.7246, 3.537, 0.895, \\ 47.821, 1.925, 1.031)} \\
\cline{2-11} 

& ${d_T} = 1000$ m & \makecell{0.7192, 0.0108 \\  (-0.0161, 0.0161)} & \makecell{0.8186, 0.0027 \\ (15.413, 0.0651)} & --,-- & \makecell{0.7945, 0.0032 \\ (4.487, 1.029)} & \makecell{0.8861, 0.0019 \\ (3.203, 2.624, 0.824)} & \makecell{0.7885, 0.0091 \\ (157.033, 16.362)} & \makecell{0.9466, 0.0017 \\ (1.3122, 3.056, \\ 11.683)} & \makecell{0.6861, 0.0138 \\ (0.1104, 12.1934, \\ 1.1098, 4.5662, 6.1409)} & \makecell{0.9665, $4.1431\times {10^{ - 4}}$ \\ (0.2386, 2.522, 0.989, \\ 1.2534, 4.695, 8.662)} \\
\hline 

\multirow{5}{*}{\makecell{${d_L} = 100$ m, \\ ${D_a} = 6$ cm}} & ${d_T} = 50$ m & \makecell{0.0793, 0.0589 \\  (-0.105, 0.105)} & \makecell{0.6719, 0.0172 \\ (6.173, 0.202)} & \makecell{-0.3583, \\ 0.0359 \\ (120.472)} & \makecell{0.6671, 0.0186 \\ (2.603, 1.261)} & \makecell{0.6913, 0.0169 \\ (2.8681, 1.571, \\ 0.895)} & \makecell{0.3208, 0.0408 \\ (3.161, 24.629)} & \makecell{0.7126, 0.0139 \\ (1.122, 3.032, \\ 2.181)} & \makecell{0.6421, 0.0286 \\ (0.1294, 4.5832, \\ 1.0011, 3.6292, 2.1435)} & \makecell{0.8873, 0.0101 \\ (0.348, 10.0415, 1.8412, \\ 1.021, 3.901, 3.124)} \\
\cline{2-11} 

& ${d_T} = 500$ m & \makecell{0.8878, 0.0265 \\  (-0.1901, 0.1901)} & \makecell{0.9687, 0.0013 \\ (1.921, 0.493)} & \makecell{0.8040, \\ 0.0316 \\ (165.619)} & \makecell{0.9794, \\ $3.8419\times {10^{ - 4}}$ \\ (1.391, 0.921)} & \makecell{0.9096, 0.0109 \\ (10.29, 0.511, 0.121)} & \makecell{0.9446, 0.0077 \\ (5.413, 3.225)} & \makecell{0.9113, 0.0090 \\ ($1.217\times {10^{ - 19}}$, \\ 11.981, 0.1083)} & \makecell{0.9628, 0.0054 \\ (0.1083, 15.2948, \\ 1.1365, 1.4255, 1.9478)} & \makecell{0.9796, $2.9716\times {10^{ - 4}}$ \\ (0.111, 1.0712, 0.502, \\ 1.0011, 1.691, 1.689)} \\
\cline{2-11} 

& ${d_T} = 1000$ m & \makecell{0.9245, \\ $9.9097\times {10^{ - 4}}$ \\ (-0.068, 0.068)} & \makecell{0.8280, 0.0021 \\ (3.8965, 0.2661)} & --,-- & \makecell{0.6986, 0.0037 \\ (2.1415, 1.1192)} & \makecell{0.8806, 0.0015 \\ (4.9145, 1.0292, \\ 0.4714)} & \makecell{0.8979, 0.0011 \\ (70.021, 4.287)} & \makecell{0.9316, $7.0397\times {10^{ - 4}}$ \\ (0.1501, 4.0721, \\ 0.862)} & \makecell{0.9424, $6.5541\times {10^{ - 4}}$ \\ (0.1009, 9.3958, \\ 0.5180, 3.2772, 1.4839)} & \makecell{0.9580, $3.0425\times {10^{ - 4}}$ \\ (0.3341, 3.570, 0.797, \\ 1.7034, 1.6301, 3.022)} \\
\hline 

\multirow{5}{*}{\makecell{${d_L} = 10$ m, \\ ${D_a} = 10$ cm}} & ${d_T} = 50$ m & \makecell{0.8031, 0.0063 \\  (-0.016, 0.016)} & \makecell{0.8099, 0.0045 \\ (15.191, 0.0632)} & --,-- & \makecell{0.8973, 0.0020 \\ (4.570, 1.024)} & \makecell{0.9023, 0.0019 \\ (2.219, 2.812, \\ 0.895)} & \makecell{0.8290, 0.0030 \\ (127.521, 18.594)} & \makecell{0.8861, 0.0023 \\ (1.2296, 3.2263, \\ 8.849)} & \makecell{0.9354, 0.0016 \\ (0.1461, 1.5079, \\ 1.0853, 4.7053, 5.0594)} & \makecell{0.9786, $3.7158\times {10^{ - 4}}$ \\ (0.6418, 5.261, 1.0261, \\ 1.3821, 2.218, 3.152)} \\
\cline{2-11} 

& ${d_T} = 500$ m & \makecell{0.4558, 0.0225 \\  (-0.018, 0.018)} & \makecell{0.7673, 0.0090 \\ (15.3028, 0.0659)} & \makecell{0.1814, \\ 0.0302 \\ (170.34)} & \makecell{0.4945, 0.0175 \\ (4.4218, 1.0469)} & \makecell{0.7715, 0.0076 \\ (12.996, 1.457, \\ 0.471)} & \makecell{0.7890, 0.0052 \\ (28.815, 29.001)} & \makecell{0.9305, 0.0044 \\ (0.1302, 12.2361, \\ 1.1694)} & \makecell{0.9478, 0.0016 \\ (0.0563, 10.7682, \\ 1.0294, 4.8060, 4.3256)} & \makecell{0.9823, $1.7789\times {10^{ - 4}}$ \\ (0.9197, 4.5985, 1.0099, \\ 0.004, 0.0821, 12.1214)} \\
\cline{2-11} 

& ${d_T} = 1000$ m & \makecell{0.8988, 0.0051 \\  (-0.0158, 0.0158)} & \makecell{0.9371, 0.0029 \\ (16.1346, 0.0625)} & --,-- & \makecell{0.9710, 0.0017 \\ (4.6223, 1.095)} & \makecell{0.9533, 0.0021 \\ (3.2394, 2.497, \\ 0.807)} & \makecell{0.9267, 0.0036 \\ (149.021, 16.921)} & \makecell{0.9454, 0.0022 \\ (0.2001, 12.2359, \\ 1.3504)} & \makecell{0.7772, 0.0086 \\ (0.1511, 8.6503, \\ 1.1451, 3.1132, 15.8190)} & \makecell{0.9870, $5.1204\times {10^{ - 4}}$ \\ (0.8023, 5.0538, 0.9972, \\ 1.004, 16.0821, 5.1214)} \\
\hline 

\multirow{5}{*}{\makecell{${d_L} = 100$ m, \\ ${D_a} = 10$ cm}} & ${d_T} = 50$ m & \makecell{0.9217, 0.0043 \\ (-0.0687, 0.0687} & \makecell{0.9347, 0.0024 \\ (4.302, 0.2361)} & \makecell{0.1144, \\ 0.0153 \\ (170.34)} & \makecell{0.9592, \\ $5.7291\times {10^{ - 4}}$ \\ (2.2698, 1.006)} & \makecell{0.9555, 0.0011 \\ (2.0553, 1.5421, \\ 0.781)} & \makecell{0.9362, 0.0023 \\ (38.766, 5.187)} & \makecell{0.9498, 0.0018 \\ (0.2689, 4.061, \\ 1.0679)} & \makecell{0.9510, 0.0016 \\ (0.2547, 1.4831, \\ 0.8142, 3.4323, 2.1005)} & \makecell{0.9666, $4.1310\times {10^{ - 4}}$ \\ (0.348, 1.515, 1.1412, \\ 1.019, 3.017, 3.1219)} \\
\cline{2-11} 

& ${d_T} = 500$ m & \makecell{0.9300, 0.0014 \\  (-0.0301, 0.0301)} & \makecell{0.9548, \\ $4.9456\times {10^{ - 4}}$ \\ (8.8902, 0.1127)} & --,-- & \makecell{0.9557, \\ $3.6453\times {10^{ - 4}}$ \\ (3.3801, 1.0201)} & \makecell{0.8714, 0.0073 \\ (3.741, 1.9741, \\ 0.782)} & \makecell{0.9521, \\ $5.0780\times {10^{ - 4}}$ \\ (170.021, 9.287)} & \makecell{0.9244, 0.0025 \\ (0.3319, 6.8605, \\ 1.3712)} & \makecell{0.9037, 0.0039 \\ (0.1422, 2.6112, \\ 0.9911, 4.6190, 3.1866)} & \makecell{0.9871, $ 2.9112\times {10^{ - 4}}$ \\ (0.8249, 3.970, 1.0397, \\ 4.034, 1.2653, 5.1084)} \\
\cline{2-11} 

& ${d_T} = 1000$ m & \makecell{0.9157, 0.0035 \\  (-0.038, -.038)} & \makecell{0.9130, 0.0041 \\ (6.9312, 0.1521)} & --,-- & \makecell{0.8941, 0.0060 \\ (3.0812, 1.1211)} & \makecell{0.9261, 0.0022 \\ (2.0134, 2.151, \\ 0.902)} & \makecell{0.9229, 0.0027 \\ (169.981, 7.262)} &  \makecell{0.7751, 0.0091 \\ (1.2998, 2.8620, \\ 3.989)} & \makecell{0.9007, 0.0052 \\ (0.2032, 4.0273, \\ 0.9097, 4.9291, 2.8477)} & \makecell{0.9804, $7.2381\times {10^{ - 4}}$ \\ (0.8012, 3.970, 1.0394, \\ 45.910, 1.955, 1.129)} \\
\hline 
\end{tabular}}
\end{center}
\vspace{-5em}
\end{table}

Based on the parameters of Table \ref{tab4}, Fig. \ref{fig9} illustrates histograms of the simulation data together with the WGG distribution as well as the existing distributions for the Indian Ocean. Similar to Fig. \ref{fig8}, it is evident that the normalized received intensity distribution with a lager aperture (\ref{fig9}(g)-(l)) is more concentrated than that with a smaller aperture (\ref{fig9}(a)-(f)). Additionally, for large-aperture conditions, the depth of transmitter placement has little effect on the distribution of normalized received intensity, while for small-aperture conditions, the impact of ocean stratification on light intensity fluctuations is significantly different. With other propagation conditions being equal, the light intensity fluctuations are more pronounced as the propagation depth increases, and the corresponding statistical histogram becomes more divergent, which can be seen from the horizontal coordinates of the subplots. Note that for the links with the large receiver aperture, except for the K distribution, which may fail in some situations, the other types of statistical models can fairly predict the fading distribution in all the vertical turbulence channel conditions. While for the links with the small receiver aperture, in the vast majority of cases, these distributions seem to fit only a portion of the data and do not follow the stretched shape of the graph very well. Surprisingly, the mixture WGG distribution yields an excellent match to the acquired simulation data for all the considered channel conditions. The results of goodness of fit tests presented in Table \ref{tab4} strongly support the application of the WGG model for turbulence induced-fading in vertical UWOC channels. 

\begin{figure}[!htbp]
\vspace{0em}
\setlength{\abovecaptionskip}{-0.3cm}
\centering
\includegraphics[width=5.5in]{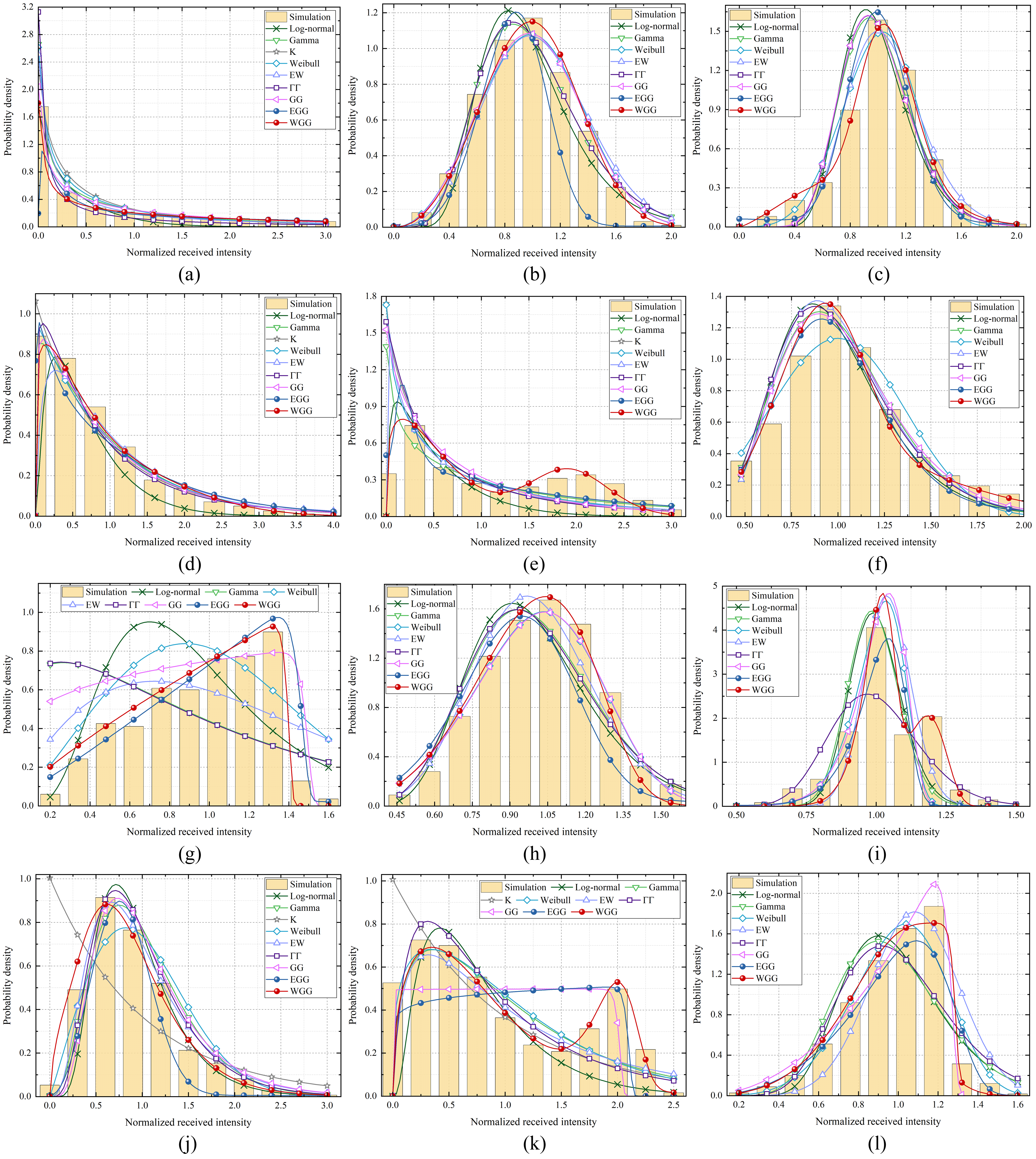}
\caption{Accordance of different statistical distributions, mentioned in section III, with the PDF histograms of the acquired simulation data through various UWOC channel conditions in Atlantic Ocean.}
\label{fig10}
\vspace{-3em}
\end{figure}
\begin{table}
\begin{center}
\caption{GoF and the constant parameters of different PDFs for the various channel conditions considered in Pacific Ocean.}
\label{tab5}
\resizebox{15cm}{5cm}{
\renewcommand\arraystretch{1.5}
\begin{tabular}{|| c || c || c || c || c || c || c || c || c || c || c ||}
\hline

\multicolumn{2}{|| c ||}{\multirow{3}{*}{\textbf{Channel Condition}}} & \textbf{Log-normal} & \textbf{Gamma} & \textbf{K} & \textbf{Weibull} & \textbf{EW} & \textbf{$\Gamma \Gamma $} & \textbf{GG} & \textbf{EGG} & \textbf{WGG} \\
\cline{3-11} 

\multicolumn{2}{|| c ||}{} & ${R^2}$, MSE & ${R^2}$, MSE & ${R^2}$, MSE & ${R^2}$, MSE & ${R^2}$, MSE & ${R^2}$, MSE & ${R^2}$, MSE & ${R^2}$, MSE & ${R^2}$, MSE \\

\multicolumn{2}{|| c ||}{} & (${\mu _X},\sigma _X^2$) & ($\tilde k,\theta $) & ($\alpha $) & ($\beta ,\tilde \eta $) & ($\alpha ,\beta ,\tilde \eta $) & ($\alpha ,\beta $) & ($\tilde a,\tilde d,\tilde p$) & ($\tilde \omega ,\tilde \lambda ,\tilde a,\tilde d,\tilde p$) & ($\varpi ,\beta ,\tilde \eta ,\tilde a,\tilde d,\tilde p$) \\ 
\hline

\multirow{5}{*}{\makecell{${d_L} = 10$ m, \\ ${D_a} = 6$ cm}} & ${d_T} = 50$ m & \makecell{0.9222, 0.0015 \\  (-1.023, 1.023)} & \makecell{0.8974, 0.0022 \\ (0.4975, 2.0325)} & --,-- & \makecell{0.9403, 0.0012 \\ (0.672, 0.758)} & \makecell{0.8783, 0.0025 \\ (3.499, 0.307, 0.102)} & \makecell{0.9178, 0.0020 \\ (0.207, 2.732)}  & \makecell{0.9488, 0.0010 \\ (2.744, 0.535, \\ 1.0102)} & \makecell{0.9247, 0.0013 \\ (0.7264, 3.7357, 3.7357, \\ 2.1345, 0.3563)} & \makecell{0.9846, $5.7610\times {10^{ - 5}}$ \\ (0.508, 0.382, 0.998, \\ 2.573, 0.934, 0.989)} \\
\cline{2-11} 

& ${d_T} = 500$ m & \makecell{0.9132, 0.0091 \\  (-0.0349, 0.0349)} & \makecell{0.9642, \\ $9.4154\times {10^{ - 4}}$ \\ (7.1431, 0.1397)} & --,-- & \makecell{0.9774, \\ $7.3324\times {10^{ - 4}}$ \\ (3.101, 1.1203)} & \makecell{0.9577, 0.0031 \\ (1.502, 2.479, 1.021)} & \makecell{0.9633, 0.0021 \\ (7.579, 170.092)} & \makecell{0.9822, $6.1357\times {10^{ - 4}}$ \\ (1.0747, 3.2051, \\ 2.9219)} & \makecell{0.6765, 0.0311 \\ (0.3102, 48.3782, 0.9147, \\ 4.4941, 4.0557)} & \makecell{0.9872, $3.4999\times {10^{ - 4}}$ \\ (0.8154, 3.1538, 1.073, \\ 1.013, 5.103, 3.082)} \\
\cline{2-11} 

& ${d_T} = 1000$ m & \makecell{0.8462, 0.0318 \\  (-0.0162, 0.0162)} & \makecell{0.9016, 0.0074 \\ (15.3792, 0.0652)} & --,-- & \makecell{0.9688, \\ $2.3383\times {10^{ - 4}}$ \\ (4.3403, 1.0987)} & \makecell{0.9680, \\ $8.7533\times {10^{ - 4}}$ \\ (1.0987, 2.115, 0.7591)} & \makecell{0.8869, 0.0110 \\ (149.01, 17.193)} & \makecell{0.8813, 0.0211 \\ (0.0441, 16.627, \\ 0.927)} & \makecell{0.9566, 0.0016 \\ (0.1219, 1.9795, 0.5125, \\ 10.2441, 2.1917)} & \makecell{0.9855, $1.9072\times {10^{ - 4}}$ \\ (0.4023, 2.274, 1.0249, \\ 0.0391, 30.233, 1.0194)} \\
\hline 

\multirow{5}{*}{\makecell{${d_L} = 100$ m, \\ ${D_a} = 6$ cm}} & ${d_T} = 50$ m & \makecell{0.9030, 0.0443 \\  (-0.3231, 0.3231)} & \makecell{0.9652, 0.0055 \\ (1.0749, 0.927)} & \makecell{0.9644, \\ 0.0084 \\ (17.83)} & \makecell{0.9609, 0.0091 \\ (1.0119, 1.0049)} & \makecell{0.9404, 0.0104 \\ (2.505, 0.6997, \\ 0.5027)} & \makecell{0.9782, 0.0041 \\ (5.011, 1.1797)} & \makecell{0.9732, 0.0051 \\ (0.6298, 1.1821, \\ 0.8438)} & \makecell{0.9240, 0.0118 \\ (0.1636, 0.2129, 0.9354, \\ 1.3414, 1.0007)} & \makecell{0.9802, 0.0010 \\ (0.3034, 2.123, 1.818, \\ 0.7118, 1.1302, 1.288)} \\
\cline{2-11} 

& ${d_T} = 500$ m & \makecell{0.3943, 0.0472 \\  (-0.5826, 0.5826)} & \makecell{0.6835, 0.0167 \\ (0.726, 1.963)} & \makecell{0.6954, \\ 0.0093 \\ (2.481)} & \makecell{0.7315, 0.0081 \\ (0.821, 0.983)} & \makecell{0.7351, 0.0070 \\ (1.487, 0.6129, \\ 0.6278)} & \makecell{0.6857, 0.0107 \\ (1.7114, 1.2164)} & \makecell{0.6753, 0.0260 \\ (0.7527, 0.9236, \\ 0.8534)} & \makecell{0.8366, 0.0050 \\ (0.8573, 1.7132, $3.9804\times {10^{ - 5}}$ \\ 11.6650, 0.3943)} & \makecell{0.8621, $4.6798\times {10^{ - 4}}$ \\ (0.6273, 1.2692, 0.582, \\ 1.024, 10.792, 2.301)} \\
\cline{2-11} 

& ${d_T} = 1000$ m & \makecell{0.9367, 0.0024 \\ (-0.0265, 0.0265)} & \makecell{0.9517, \\ $9.0813\times {10^{ - 4}}$ \\ (9.7997, 0.1023)} & --,-- & \makecell{0.8730, 0.0124 \\ (3.266, 1.118)} & \makecell{0.9438, 0.0018 \\ (9.118, 1.358, \\ 0.478)} & \makecell{0.9439, 0.0016 \\ (19.601, 19.421)} & \makecell{0.9534, $9.0409\times {10^{ - 4}}$ \\ ($3.811\times {10^{ - 4}}$, \\ 19.8192, 0.4742)} & \makecell{0.9706, $4.8611\times {10^{ - 4}}$ \\ (0.2175, 0.9782, 0.1161, \\ 11.7730, 1.1029)} & \makecell{0.9869, $4.1294\times {10^{ - 5}}$ \\ (0.6186, 4.362, 0.961, \\ $7.125\times {10^{ - 4}}$, 17.727, 0.4803)} \\
\hline 

\multirow{5}{*}{\makecell{${d_L} = 10$ m, \\ ${D_a} = 10$ cm}} & ${d_T} = 50$ m & \makecell{-0.1397, 0.0405 \\  (-0.0711, 0.0711)} & \makecell{-0.4787, 0.0083 \\ (1.3613, 0.7321)} & --,-- & \makecell{0.3400, 0.0174 \\ (2.3009, 1.1289)} & \makecell{0.2435, 0.0212 \\ (1.232, 1.518, \\ 1.120)} & \makecell{-0.4923, 0.0073 \\ (1.348, 171.352)} & \makecell{0.4568, 0.0124 \\ (1.498, 1.203, \\ 53.8101)} & \makecell{0.7044, 0.0012 \\ (0.2639, 10.1349, 1.4691, \\ 2.0862, 41.9828)} & \makecell{0.8902, $9.4894\times {10^{ - 5}}$ \\ (0.249, 12.422, 0.142, \\ 1.402, 1.8131, 73.891)} \\
\cline{2-11} 

& ${d_T} = 500$ m & \makecell{0.8855, 0.0027 \\  (-0.0167, 0.0167)} & \makecell{0.9323, 0.0015 \\ (15.127, 0.0662)} & --,-- & \makecell{0.9655, \\ $9.1955\times {10^{ - 4}}$ \\ (4.577, 1.0951)} & \makecell{0.9692, \\ $8.2009\times {10^{ - 4}}$ \\ (3.089, 2.689, 0.828)} & \makecell{0.9218, 0.0019 \\ (155.303, 16.478)} & \makecell{0.9632, $9.9467\times {10^{ - 4}}$ \\ (1.101, 4.549, \\ 4.6212)} & \makecell{0.8290, 0.0062 \\ (0.1812, 2.8361, 0.9526, \\ 5.2644, 4.3170)} & \makecell{0.9834, $ 9.2225\times {10^{ - 5}}$ \\ (0.8026, 4.681, 1.032, \\ 1.179, 8.251, 7.783)} \\
\cline{2-11} 

& ${d_T} = 1000$ m & \makecell{0.5395, 0.0174 \\  (-0.0021, 0.0021)} & \makecell{0.4802, 0.0290 \\ (120.992, 0.0082)} & --,-- & \makecell{0.5582, 0.0124 \\ (13.173, 1.0411)} & \makecell{0.6389, 0.0077 \\ (7.293, 5.6892, \\ 0.891)} & \makecell{0.5879, 0.0108 \\ (78.322, 79.128)} & \makecell{0.5194, 0.0260 \\ (1.0721, 12.012, \\ 16.812)} & \makecell{0.6114, 0.0095 \\ (0.2255, 0.2255, 1.0721, \\ 11.9117, 17.5004)} & \makecell{0.9428, 0.0031 \\ (0.5893, 20.311, 1.0213, \\ 1.194, 15.612, 16.926)} \\
\hline 

\multirow{5}{*}{\makecell{${d_L} = 100$ m, \\ ${D_a} = 10$ cm}} & ${d_T} = 50$ m & \makecell{0.8641, 0.0078 \\ (-0.0656, 0.0656)} & \makecell{0.9068, 0.0022 \\ (3.9221, 0.258)} & \makecell{0.4956, \\ 0.0425 \\ (171.6201)} & \makecell{0.8936, 0.0025 \\ (2.0651, 1.1303)} & \makecell{0.9555, 0.0012 \\ (5.029, 0.928, \\ 0.4101)} & \makecell{0.9148, 0.0016 \\ (7.778, 7.932)} & \makecell{0.8744, 0.0035 \\ ($1.78\times {10^{ - 7}}$, \\ 14.318, 0.2588)} & \makecell{0.8154, 0.0156 \\ (0.3132, 56.2457, \\ 0.8383, 3.035, 2.6967)} & \makecell{0.9807, $3.0154\times {10^{ - 4}}$ \\ (0.601, 1.6189, 0.821, \\ $2.13\times {10^{ - 7}}$, 14.803, 0.2631)} \\
\cline{2-11} 

& ${d_T} = 500$ m & \makecell{0.1548, 0.0198 \\  (-0.2051, 0.2051)} & \makecell{0.5536, 0.0038 \\ (1.5003, 0.711)} & \makecell{0.4593, \\ 0.0076 \\ (171.618)} & \makecell{0.5445, 0.0050 \\ (1.291, 1.0798)} & \makecell{0.5560, 0.0032 \\ (2.497, 0.7189, \\ 0.5589)} & \makecell{0.5263, 0.0063 \\ (5.023, 1.917)} & \makecell{0.0935, 0.0262 \\ (2.0268, 1.002, \\ 72.989)} & \makecell{0.0583, 0.0403 \\ (0.0104, 51.2196, 2.1062, \\ 1.0785, 70.8761)} & \makecell{0.9582, 0.0018 \\ (0.771, 1.4189, 0.821, \\ 2.0351, 11.129, 11.692)} \\
\cline{2-11} 

& ${d_T} = 1000$ m & \makecell{0.7349, 0.0189 \\  (-0.0183, 0.0183)} & \makecell{0.7372, 0.0160 \\ (12.989, 0.0749)} & --,-- & \makecell{0.9101, 0.0046 \\ (4.911, 1.088)} & \makecell{0.8501, 0.0118 \\ (2.051, 3.811, \\ 1.019)} & \makecell{0.7708, 0.0145 \\ (13.812, 161.461)} &  \makecell{0.9040, 0.0053 \\ (1.2701, 3.102, \\ 44.512)} & \makecell{0.9003, 0.0071 \\ (0.1013, 29.3893, \\ 1.2558, 3.5066, 9.5560)} & \makecell{0.9299, 0.0012 \\ (0.3892, 5.028, 1.011, \\ 1.289, 3.271, 74.022)} \\
\hline 
\end{tabular}}
\end{center}
\vspace{-5em}
\end{table}

For Atlantic Ocean environment, Fig. \ref{fig10} presents the accordance of different statistical distributions with the PDF histograms of the acquired simulation data. Furthermore, detailed simulation results for GoF values of each of the considered statistical distributions and their corresponding constant coefficients are listed in Table \ref{tab5}. Comparing Figs. \ref{fig10}(a)-(f) and Figs. \ref{fig10}(g)-(l), it is observed that increasing the aperture size can distinctly reduce the intensity fluctuations due to the aperture averaging effect. Keeping other propagation conditions the same, the increasing propagation depth may cause the strong fluctuations in light intensity so that the corresponding statistical histogram becomes more divergent. One particular phenomenon is that the histograms corresponding to ${d_T} = 500$ m, ${d_L} = 100$ m, ${D_a} = 6$ cm and ${d_T} = 500$ m, ${d_L} = 100$ m, ${D_a} = 10$ cm occur peaks on both sides of the mean of the acquired data. From Fig. \ref{fig6}, we note that the temperature and salinity remain stable near a depth of 600 m. The occurrence of constant temperature and constant salinity layer in the thermocline signifies that there is strong mixing of seawater in this depth, generating the stronger turbulence, which exacerbates the fading deterioration effect and causes the received intensity to mainly lie either in large or small values. In such circumstances, the typical single-lobe distributions cannot appropriately fit the acquired data and generally a two-lobe statistical distribution is required to predict the statistical behavior of vertical UWOC fading in all of the considered channel conditions. However, exploring Figs. \ref{fig10}(e), (k) and the fifth and eleventh rows of Table \ref{tab5}, the mixed EGG distribution obtained by the EM algorithm optimization is closer to the single-lobe GG distribution, which also does not fit well to double-peaked histogram. Fortunately, the unified WGG model always fits very well to the acquired data. From Table \ref{tab5}, it is evident that ${R^2}$ values corresponding to the WGG model are the highest and the MSE measures associated with the WGG model have the smallest values. Overall, the WGG distribution gives the best performance in terms of quality of fit to the acquired data. 

\section{Error Rate Performance of Underwater Vertical Links}
Given the ability of the WGG distribution to fit well with the obtained data under various channel conditions, we use this PDF to formulate a closed form expression for BER performance further investigate the effect of channel parameters on BER performance of vertical UWOC systems.

\subsection{Analytical formulae of the average BER}
We consider a vertical UWOC system with intensity modulation and direct detection. Using the On-Off Keying modulation deployment, the conditional BER takes the form of 
\begin{equation}
\label{ex57}
BER = \frac{1}{2}{\rm{erfc}}\left( {\frac{{\tilde \gamma  \cdot I}}{{2\sqrt 2 }}} \right),
\end{equation}
where $\tilde \gamma $ denotes the SNR in the absence of turbulence \cite{ref52}. In \eqref{ex57}, ${\rm{erfc}}\left(  \cdot  \right)$ is the complementary error function. The average BER can be calculated by averaging the conditional BER over the PDF of $I$ as
\begin{equation}
\label{ex58}
\left\langle {BER} \right\rangle  = \frac{1}{2}\int\limits_0^\infty  {{f_I}\left( I \right){\rm{erfc}}\left( {\frac{{\tilde \gamma  \cdot I}}{{2\sqrt 2 }}} \right)} dI.
\end{equation}
Replacing \eqref{ex43} and \eqref{ex44} in \eqref{ex58}, we have 
\begin{equation}
\label{ex59}
\left\langle {BER} \right\rangle  = \frac{{\varpi \beta }}{{2{{\tilde \eta }^\beta }}}\int\limits_0^\infty  {{I^{\beta  - 1}}{e^{ - {{\left( {\frac{I}{{\tilde \eta }}} \right)}^\beta }}}erfc\left( {\frac{{\tilde \gamma  \cdot I}}{{2\sqrt 2 }}} \right)} dI + \frac{{\left( {1 - \varpi } \right)\tilde p}}{{2{{\tilde a}^{\tilde d}}\Gamma \left( {\tilde d/\tilde p} \right)}}\int\limits_0^\infty  {{I^{\tilde d - 1}}{e^{ - {{\left( {\frac{I}{{\tilde a}}} \right)}^{\tilde p}}}}erfc\left( {\frac{{\tilde \gamma  \cdot I}}{{2\sqrt 2 }}} \right)} dI.
\end{equation}
By using the equation provided in \cite{ref49} (Eqs. (2.8-5) and (2.8-6)), 
\begin{equation}
\int\limits_0^\infty  {{x^{{a_1} - 1}}{e^{ - {a_2}{x^{{a_3}}}}}{\rm{erfc}}\left( {{a_4}x} \right)} dx = \left\{ \begin{array}{l}
\frac{{ - {a_4}}}{{\sqrt \pi  {a_3}a_2^{\left( {{a_1} + 1} \right)/{a_3}}}}\sum\limits_{t = 0}^\infty  {\frac{{{{\left( { - 1} \right)}^t}}}{{\left( {t + 1/2} \right)t!}}} \Gamma \left( {\frac{{2t + {a_1} + 1}}{{{a_3}}}} \right)\frac{{a_4^{2t}}}{{a_2^{\frac{{2t}}{{{a_3}}}}}} + \frac{{\Gamma \left( {{a_1}/{a_3}} \right)}}{{{a_3}a_2^{{a_1}/{a_3}}}},{a_3} > 2,\\
\frac{1}{{a_4^{{a_1}}\sqrt \pi  }}\sum\limits_{t = 0}^\infty  {\frac{1}{{\left( {{a_3}t + {a_1}} \right)t!}}} \Gamma \left( {\frac{{{a_3}t + {a_1} + 1}}{2}} \right){\left( {\frac{{ - {a_2}}}{{a_4^{{a_3}}}}} \right)^t},{a_3} < 2,\\
\frac{{ - {a_4}}}{{\sqrt \pi  a_2^{\left( {\alpha  + 1} \right)/2}}}\Gamma \left( {\frac{{{a_1} + 1}}{2}} \right){}_2{F_1}\left( {\frac{1}{2},\frac{{{a_1} + 1}}{2};\frac{3}{2}; - \frac{{a_4^2}}{{{a_2}}}} \right) + \frac{{\Gamma \left( {{a_1}/2} \right)}}{{2a_2^{{a_1}/2}}},{a_3} = 2,
\end{array} \right.
\end{equation}
the integral form in \eqref{ex59} can be simplified. Following equation is for the first term of \eqref{ex59}, 
\begin{equation}
{\left\langle {BER} \right\rangle _1} = \left\{ \begin{array}{l}
\frac{{ - \varpi \tilde \gamma \tilde \eta }}{{4\sqrt {2\pi } }}\sum\limits_{t = 0}^\infty  {\frac{{{{\left( { - 1} \right)}^t}}}{{\left( {t + 1/2} \right)t!}}} \Gamma \left( {\frac{{2t + \beta  + 1}}{\beta }} \right){\left( {\frac{{\tilde \gamma \tilde \eta }}{{2\sqrt 2 }}} \right)^{2t}} + \frac{\varpi }{2},\left( {\beta  > 2} \right),\\
\frac{{\varpi \beta }}{{2\sqrt \pi  }}{\left( {\frac{{2\sqrt 2 }}{{\tilde \gamma \tilde \eta }}} \right)^\beta }\sum\limits_{t = 0}^\infty  {\frac{{\Gamma \left( {\left( {\beta t + \beta  + 1} \right)/2} \right)}}{{\left( {\beta t + \beta } \right)t!}}} {\left( { - {{\left( {\frac{{2\sqrt 2 }}{{\tilde \gamma \tilde \eta }}} \right)}^\beta }} \right)^t},\left( {\beta  < 2} \right),\\
\frac{{ - \tilde \gamma \tilde \eta \varpi }}{{2\sqrt {2\pi } }}\Gamma \left( {\frac{3}{2}} \right){}_2{F_1}\left( {\frac{1}{2},\frac{3}{2};\frac{3}{2}; - \frac{{{{\tilde \gamma }^2}}}{{8{a_2}}}} \right) + \frac{\varpi }{2},\left( {\beta  = 2} \right),
\end{array} \right.
\end{equation}
and for the second term, 
\begin{equation}
{\left\langle {BER} \right\rangle _2} = \left\{ \begin{array}{l}
\frac{{\left( {\varpi  - 1} \right)}}{{\Gamma \left( {\tilde d/\tilde p} \right)}}\left( {\frac{{\tilde a\tilde \gamma }}{{4\sqrt {2\pi } }}} \right)\sum\limits_{t = 0}^\infty  {\frac{{{{\left( { - 1} \right)}^t}\Gamma \left( {\left( {2t + \tilde d + 1} \right)/\tilde p} \right)}}{{\left( {t + 1/2} \right)t!}}} {\left( {\frac{{\tilde a\tilde \gamma }}{{2\sqrt 2 }}} \right)^{2t}} + \frac{{1 - \varpi }}{2},\left( {\tilde p > 2} \right),\\
\frac{{\left( {1 - \varpi } \right)\tilde p}}{{2{{\tilde a}^{\tilde d}}\Gamma \left( {\tilde d/\tilde p} \right)\sqrt \pi  }}{\left( {\frac{{2\sqrt 2 }}{{\tilde \gamma }}} \right)^{\tilde d}}\sum\limits_{t = 0}^\infty  {\frac{{\Gamma \left( {\left( {\tilde pt + \tilde d + 1} \right)/2} \right)}}{{\left( {\tilde pt + \tilde d} \right)t!}}} {\left( { - {{\left( {\frac{{2\sqrt 2 }}{{\tilde a\tilde \gamma }}} \right)}^{\tilde p}}} \right)^t},\left( {\tilde p < 2} \right),\\
\frac{{\left( {1 - \varpi } \right)}}{{\Gamma \left( {\tilde d/2} \right)}}\frac{{ - \tilde \gamma \tilde a}}{{2\sqrt {2\pi } }}\Gamma \left( {\frac{{\tilde d + 1}}{2}} \right){}_2{F_1}\left( {\frac{1}{2},\frac{{\tilde d + 1}}{2};\frac{3}{2}; - \frac{{{{\left( {\tilde a\tilde \gamma } \right)}^2}}}{8}} \right) + \frac{{\left( {1 - \varpi } \right)\tilde p}}{4},\left( {\tilde p = 2} \right).
\end{array} \right.
\end{equation}
Ultimately, based on the WGG distribution parameters $\beta $ and $\tilde p$ obtained by fitting the simulation data, we can determine the closed form of $\left\langle {BER} \right\rangle $ by calculating $\left\langle {BER} \right\rangle  = {\left\langle {BER} \right\rangle _1} + {\left\langle {BER} \right\rangle _2}$.

\subsection{Performance analysis of optical communication system for oceanic vertical channel}
Based on the model of the vertical channel for optical communication established in section II, the system performance simulation is performed by using the seawater data in Fig. \ref{fig6}, measured by Argo in several ocean areas. And the main parameters used in the simulations are listed in Table \ref{tab2}. 

\begin{figure}[!htbp]
\vspace{-0em}
\setlength{\abovecaptionskip}{-0.3cm}
\centering
\includegraphics[width=6in]{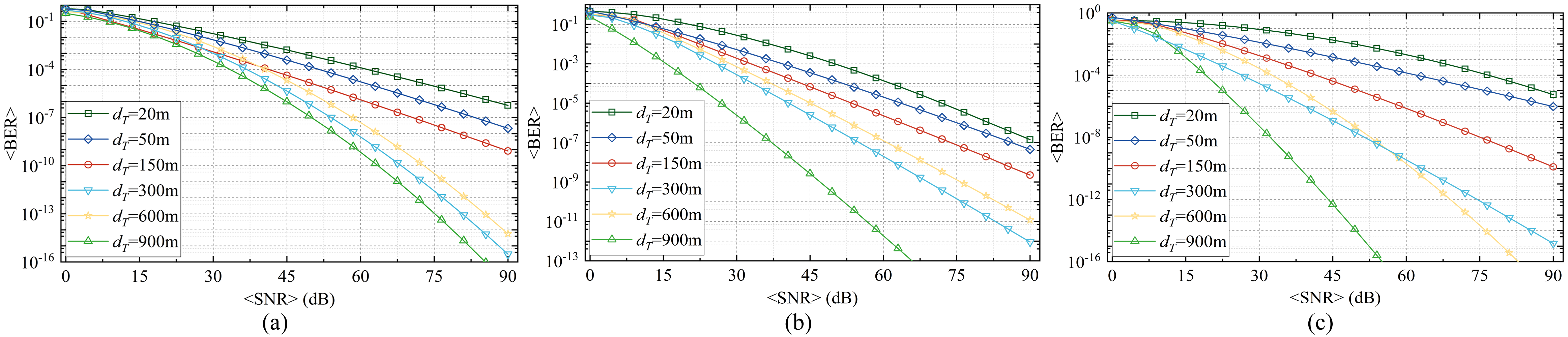}
\caption{Average BERs for transmitters at different depths in (a) Pacific, (b) Indian, (c) Atlantic Ocean.}
\label{fig11}
\vspace{-2.5em}
\end{figure}
Setting ${d_L} = 70m$, the BER performance with different optical transmitter depths in three ocean areas is shown in Fig. \ref{fig11}. It is obvious that the BERs decrease with an increase of the average SNR, and the initial depth of the optical transmitter can significantly affect the communication quality. For instance, the system BERs are the highest when the transmitters are located at a depth of 20 m, which are close to the surface of the mixed layer with strong turbulence strength. Comparing the system BER curves under different depths of the transmitters, it is observed that the BER corresponding to the deep layer (900m) is minimal, and the system BERs under different depths of transmitters in the thermocline (300,600m) are generally lower than those in the mixed layer (20m, 50m, 150m). However, in the thermocline, there may be fluctuations in temperature or salinity at some depths, which are influenced by factors such as dark currents, and probably cause the stronger turbulence. Consequently, when the average SNR is not large enough, there may be a situation where the BER at a depth of 600m is higher than that at 300m. 

\begin{figure}[!htbp]
\vspace{-1.5em}
\setlength{\abovecaptionskip}{-0.3cm}
\centering
\includegraphics[width=6in]{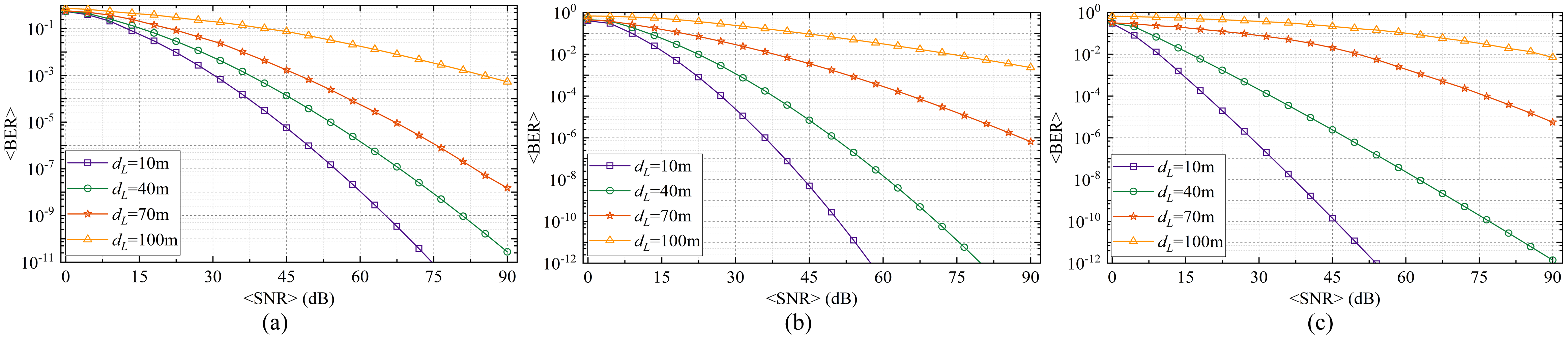}
\caption{Average BERs for different propagation depths in (a) Pacific, (b) Indian, (c) Atlantic Ocean.}
\label{fig12}
\vspace{-2em}
\end{figure}
Fig. \ref{fig12} presents the variation of average BER with average SNR at different propagation depths. It is clear that the average BER decreases with the increasing of the average SNR. Additionally, as the depth of propagation increases, the average BER performance exhibits a gradual decline. Since the transmitter depth is set to ${d_T} = 20m$, thus, the system positions corresponding to these four propagation depths are in the mixed layer, where the turbulence strength is positively related to the propagation path length, which makes the BER performance decrease with the increase of propagation depth. 

\begin{figure}[!htbp]
\vspace{0em}
\setlength{\abovecaptionskip}{-0.3cm}
\centering
\includegraphics[width=6in]{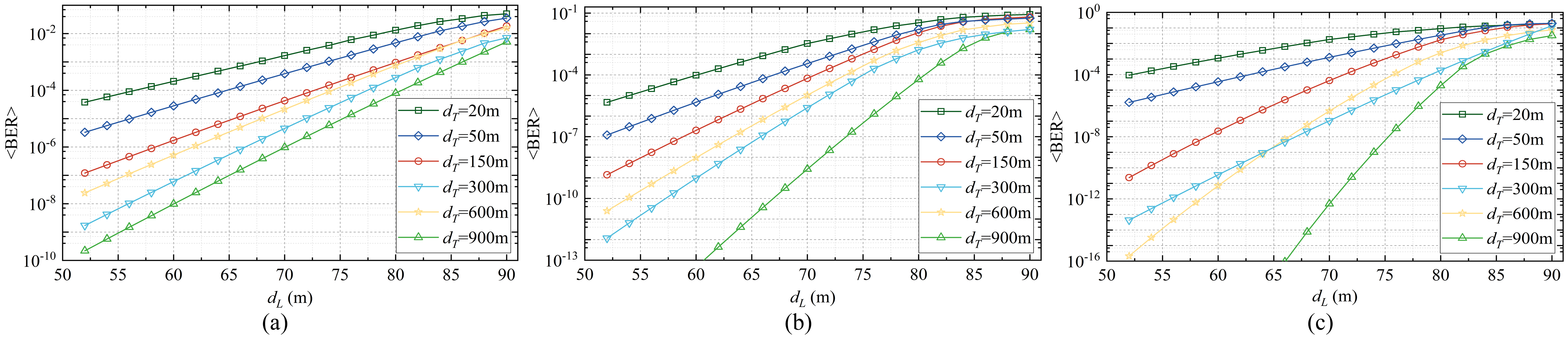}
\caption{Average BERs for transmitters at different depths with distances in (a) Pacific, (b) Indian, (c) Atlantic Ocean.}
\label{fig13}
\vspace{-2em}
\end{figure}
With ${\mathop{\rm SNR}\nolimits}  = 45dB$, Fig. \ref{fig13} shows the BER performance versus link depths at different initial depths of the optical transmitter locations. Obviously, the BERs increase with an increase of the communication distance. Similar to Fig. \ref{fig11}, the system BERs under different transmitters depths in the mixed layer are generally higher than those in the thermocline. Additionally, it is observed from Fig. \ref{fig13}(c) that when the propagation depth is less than 65m, the system BER at a depth of 300m is higher than that of 600m, while when the propagation distance is greater than 65m, the situation is the opposite. This can be explained by the fact that the occurrence of constant temperature and constant salinity layers in the thermocline leads to the increase of turbulence strength, which degrades the BER performance more significantly with the increase of propagation depth. 

\begin{figure}[!htbp]
\vspace{-1.5em}
\setlength{\abovecaptionskip}{-0.3cm}
\centering
\includegraphics[width=6in]{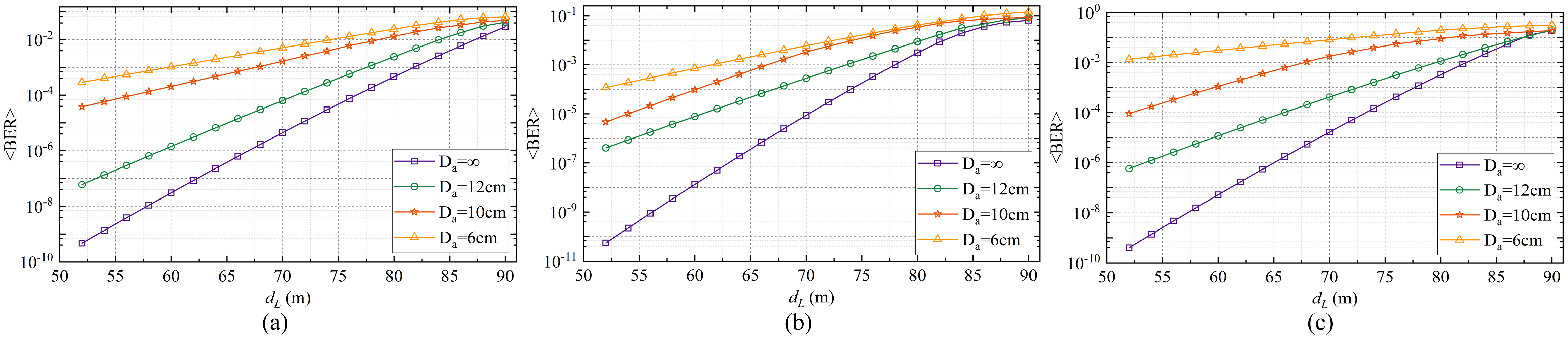}
\caption{Average BERs for different aperture size versus distances in (a) Pacific, (b) Indian, (c) Atlantic Ocean.}
\label{fig14}
\vspace{-2em}
\end{figure}
Fig. \ref{fig14} gives the average BER curves versus propagation depth for different receiver aperture size. It can be seen that the BERs increase with an extension of the link depth. In addition, the BER is significantly improved as the receiver aperture size increases. This is because an increase in aperture size can average out the turbulence-induced intensity fluctuations. Thus, in the actual communication scenario, we can consider the appropriate selection of a larger receiving aperture to achieve the purpose of improving system performance. 

\section{Conclusions and Future Directions}
In this paper, we establish a reasonable WOS-based turbulence modelling scheme for vertical UWOC channels on basis of Fourier optics theory, and provide a comprehensive study on the statistics of fading in vertical UWOC channels under various conditions. We observe that because of the aperture averaging effect, the normalized received optical intensity distribution with a lager aperture is more concentrated than that with a smaller aperture. We also note that increasing the link depth significantly exacerbates the degree of light intensity fluctuation. Additionally, the transmitter at the deep layer performs more concentrated statistical distribution compared to the transmitter at the mixed layer and thermocline.  

After the extensive simulation realizations, we propose a unified statistical model named mixture WGG model to characterize the turbulence-induced intensity fluctuations for vertical UWOC channels. We further demonstrate that this statistical model perfectly matches the data acquired under different channel conditions. Applying the WGG model, we investigate the BER performance of the vertical UWOC link under different operational conditions. Our thorough numerical studies reveal that, the system BER is distinctly degraded with an increase in the link depth and a reduction in the receiver aperture size. In general, the vertical link in the mixed layer has poorer communication quality compared to the vertical link in the thermocline. However, there may be strong seawater mixing at some depths in the thermocline, making the BER generally higher. We anticipate that our studies will catalyze the development of robust and reliable underwater communication systems and help push the frontiers of UWOC research towards the goal of seamless and high-speed underwater wireless networks.

\bibliographystyle{reference} 
\bibliography{reference}

\begin{thebibliography}{10}
\baselineskip 12pt
\providecommand{\url}[1]{#1}
\csname url@samestyle\endcsname
\providecommand{\newblock}{\relax}
\providecommand{\bibinfo}[2]{#2}
\providecommand{\BIBentrySTDinterwordspacing}{\spaceskip=0pt\relax}
\providecommand{\BIBentryALTinterwordstretchfactor}{4}
\providecommand{\BIBentryALTinterwordspacing}{\spaceskip=\fontdimen2\font plus
\BIBentryALTinterwordstretchfactor\fontdimen3\font minus
  \fontdimen4\font\relax}
\providecommand{\BIBforeignlanguage}[2]{{%
\expandafter\ifx\csname l@#1\endcsname\relax
\typeout{** WARNING: IEEEtran.bst: No hyphenation pattern has been}%
\typeout{** loaded for the language `#1'. Using the pattern for}%
\typeout{** the default language instead.}%
\else
\language=\csname l@#1\endcsname
\fi
#2}}
\providecommand{\BIBdecl}{\relax}
\BIBdecl

\bibitem{ref1}
Z.~Zeng, S.~Fu, H.~Zhang, Y.~Dong, and J.~Cheng, ``A survey of underwater
  optical wireless communications,'' \emph{IEEE Communications Surveys \&
  Tutorials}, vol.~19, no.~1, pp. 204--238, 2017.

\bibitem{ref2}
C.~Gussen, P.~Diniz, M.~Campos, W.~A. Martins, and J.~N. Gois, ``A survey of
  underwater wireless communication technologies,'' \emph{Journal of
  Communication and Information Systems}, vol.~31, no.~1, pp. 242--255, 2016.

\bibitem{MCS1}
W.~Liu, D.~Zou, Z.~Xu, and J.~Yu, ``Non-line-of-sight scattering channel
  modeling for underwater optical wireless communication,'' in \emph{2015 IEEE
  International Conference on Cyber Technology in Automation, Control, and
  Intelligent Systems (CYBER)}, 2015.

\bibitem{MCS2}
C.~Gabriel, M.-A. Khalighi, S.~Bourennane, P.~Léon, and V.~Rigaud,
  ``Monte-carlo-based channel characterization for underwater optical
  communication systems,'' \emph{Journal of Optical Communications and
  Networking}, vol.~5, no.~1, pp. 1--12, 2013.

\bibitem{MCS3}
A.~A.~B. Umar, M.~S. Leeson, and I.~Abdullahi, ``Modelling impulse response for
  nlos underwater optical wireless communications,'' in \emph{2019 15th
  International Conference on Electronics, Computer and Computation (ICECCO)},
  2019.

\bibitem{ref4}
L.~C. Andrews and R.~L. Phillips, \emph{Laser Beam Propagation through Random
  Media, Second Edition}.\hskip 1em plus 0.5em minus 0.4em\relax SPIE Press,
  Second Edition, 2005.

\bibitem{ref3}
N.~Saeed, A.~Celik, T.~Y. Al-Naffouri, and M.-S. Alouini, ``Underwater optical
  wireless communications, networking, and localization: A survey,'' \emph{Ad
  Hoc Networks}, vol.~94, p. 101935, 2018.

\bibitem{ref7}
X.~Yi, Z.~Li, and Z.~Liu, ``Underwater optical communication performance for
  laser beam propagation through weak oceanic turbulence,'' \emph{Applied
  Optics}, vol.~54, no.~6, pp. 1273--1278, 2015.

\bibitem{ref8}
M.~Sharifzadeh and M.~Ahmadirad, ``Performance analysis of underwater wireless
  optical communication systems over a wide range of optical turbulence,''
  \emph{Optics Communications}, vol. 427, pp. 609--616, 2018.

\bibitem{ref9}
Y.~Fu and Y.~Du, ``Performance of heterodyne differential phase-shift-keying
  underwater wireless optical communication systems in gamma-gamma-distributed
  turbulence,'' \emph{Appl Opt}, vol.~57, no.~9, pp. 2057--2063, 2018.

\bibitem{ref10}
G.~Xu and J.~Lai, ``Average capacity analysis of underwater optical plane wave
  over anisotropic moderate-to-strong oceanic turbulence channels with the
  málaga fading model,'' \emph{Optics Express}, vol.~28, no.~16, pp.
  24\,056--24\,068, 2020.

\bibitem{ref11}
X.~Xu, Y.~Li, P.~Huang, M.~Ju, and G.~Tan, ``Ber performance of uwoc with apd
  receiver in wide range oceanic turbulence,'' \emph{IEEE Access}, vol.~10, pp.
  25\,203--25\,218, 2022.

\bibitem{ref40}
V.~V. Nikishov and V.~I. Nikishov, ``Spectrum of turbulent fluctuations of the
  sea-water refraction index,'' \emph{International Journal of Fluid Mechanics
  Research}, vol.~27, no.~1, pp. 82--98, 2000.

\bibitem{ref12}
H.~M. Oubei, E.~Zedini, R.~T. Elafandy, A.~Kammoun, and B.~S. Ooi, ``Simple
  statistical channel model for weak temperature-induced turbulence in
  underwater wireless optical communication systems,'' \emph{Optics Letters},
  vol.~42, no.~13, pp. 2455--2458, 2017.

\bibitem{ref13}
H.~M. Oubei, E.~Zedini, R.~T. ElAfandy, A.~Kammoun, T.~K. Ng, M.-S. Alouini,
  and B.~S. Ooi, ``Efficient weibull channel model for salinity induced
  turbulent underwater wireless optical communications,'' in \emph{2017
  Opto-Electronics and Communications Conference (OECC) and Photonics Global
  Conference (PGC)}, 2017, pp. 1--2.

\bibitem{ref14}
M.~V. Jamali, A.~Mirani, A.~Parsay, B.~Abolhassani, P.~Nabavi, A.~Chizari,
  P.~Khorramshahi, S.~Abdollahramezani, and J.~A. Salehi, ``Statistical studies
  of fading in underwater wireless optical channels in the presence of air
  bubble, temperature, and salinity random variations,'' \emph{IEEE
  Transactions on Communications}, vol.~66, no.~10, pp. 4706--4723, 2018.

\bibitem{ref15}
E.~Zedini, H.~M. Oubei, A.~Kammoun, M.~Hamdi, and M.~S. Alouini, ``A new simple
  model for underwater wireless optical channels in the presence of air
  bubbles,'' in \emph{GLOBECOM 2017 - 2017 IEEE Global Communications
  Conference}, 2018.

\bibitem{ref16}
M.~V. Jamali, P.~Khorramshahi, A.~Tashakori, A.~Chizari, S.~Shahsavari,
  S.~AbdollahRamezani, M.~Fazelian, S.~Bahrani, and J.~A. Salehi, ``Statistical
  distribution of intensity fluctuations for underwater wireless optical
  channels in the presence of air bubbles,'' in \emph{2016 Iran Workshop on
  Communication and Information Theory (IWCIT)}, 2016, pp. 1--6.

\bibitem{ref17}
E.~Zedini, H.~M. Oubei, A.~Kammoun, M.~Hamdi, B.~S. Ooi, and M.-S. Alouini,
  ``Unified statistical channel model for turbulence-induced fading in
  underwater wireless optical communication systems,'' \emph{IEEE Transactions
  on Communications}, vol.~67, no.~4, pp. 2893--2907, 2019.

\bibitem{ref18}
Y.~M. Shishter, R.~Young, and F.~H. Ali, ``Modelling the intensity distribution
  of underwater optical wave propagation in the presence of turbulence and air
  bubbles,'' \emph{Optik}, vol. 270, p. 170006, 2022.

\bibitem{ref19}
A.~Ishimaru \emph{et~al.}, \emph{Wave propagation and scattering in random
  media}.\hskip 1em plus 0.5em minus 0.4em\relax Academic press New York, 1978,
  vol.~2.

\bibitem{ref20}
J.~Zhang, L.~Kou, Y.~Yang, F.~He, and Z.~Duan, ``Monte-carlo-based optical
  wireless underwater channel modeling with oceanic turbulence,'' \emph{Optics
  Communications}, vol. 475, p. 126214, 2020.

\bibitem{ref21}
Y.~M. Shishter, R.~Young, and F.~H. Ali, ``A general approach for determining
  the irradiance distribution of em waves propagating through random media,''
  \emph{IEEE Transactions on Antennas and Propagation}, vol.~70, no.~11, pp.
  10\,917--10\,924, 2022.

\bibitem{ref22}
R.~Cai, M.~Zhang, D.~Dai, Y.~Shi, and S.~Gao, ``Analysis of the underwater
  wireless optical communication channel based on a comprehensive
  multiparameter model,'' \emph{Applied Sciences}, vol.~11, no.~13, p. 6051,
  2021.

\bibitem{ref23}
M.~Elamassie and M.~Uysal, ``Performance characterization of vertical
  underwater vlc links in the presence of turbulence,'' in \emph{11th IEEE/IET
  International Symposium on Communication Systems, Networks \& Digital Signal
  Processing (CSNDSP18)}, 2018.

\bibitem{ref24}
R.~Sharma and Y.~N. Trivedi, ``Performance analysis of vertical multihop
  cooperative underwater visible light communication system with imperfect
  channel state information,'' \emph{Optical Engineering}, vol.~61, no.~4,
  pp.~--, 2022.

\bibitem{ref25}
I.~C. Ijeh, M.~A. Khalighi, M.~Elamassie, S.~Hranilovic, and M.~Uysal, ``Outage
  probability analysis of a vertical underwater wireless optical link subject
  to oceanic turbulence and pointing errors,'' \emph{Journal of Optical
  Communications and Networking}, vol.~14, no.~6, pp. 439--453, 2022.

\bibitem{ref26}
M.~Elamassie, S.~M. Sait, and M.~Uysal, ``Underwater visible light
  communications in cascaded gamma-gamma turbulence,'' in \emph{2018 IEEE
  Globecom Workshops (GC Wkshps)}, 2018, pp. 1--6.

\bibitem{ref27}
M.~Elamassie and M.~Uysal, ``Vertical underwater vlc links over cascaded
  gamma-gamma turbulence channels with pointing errors,'' in \emph{2019 IEEE
  International Black Sea Conference on Communications and Networking
  (BlackSeaCom)}, 2019, pp. 1--5.

\bibitem{ref28}
------, ``Vertical underwater visible light communication links: Channel
  modeling and performance analysis,'' \emph{IEEE Transactions on Wireless
  Communications}, vol.~19, no.~10, pp. 6948--6959, 2020.

\bibitem{ref29}
Y.~Lou, J.~Cheng, D.~Nie, and G.~Qiao, ``Performance of vertical underwater
  wireless optical communications with cascaded layered modeling,'' \emph{IEEE
  Transactions on Vehicular Technology}, vol.~71, no.~5, pp. 5651--5655, 2022.

\bibitem{ref30}
Z.~Rahman, N.~V. Tailor, S.~M. Zafaruddin, and V.~K. Chaubey, ``Unified
  performance assessment of optical wireless communication over multi-layer
  underwater channels,'' \emph{IEEE Photonics Journal}, vol.~14, no.~5, pp.
  1--14, 2022.

\bibitem{ref31}
X.~Ji, H.~Yin, L.~Jing, Y.~Liang, and J.~Wang, ``Modeling and performance
  analysis of oblique underwater optical communication links considering
  turbulence effects based on seawater depth layering,'' \emph{Opt. Express},
  vol.~30, no.~11, pp. 18\,874--18\,888, May 2022.

\bibitem{ref33}
\BIBentryALTinterwordspacing
Argo data sources. [Online]. Available:
  \url{https://argo.ucsd.edu/data/data-from-gdacs/}
\BIBentrySTDinterwordspacing

\bibitem{ref32}
J.~D. Schmidt, \emph{Numerical simulation of optical wave propagation: With
  examples in MATLAB}.\hskip 1em plus 0.5em minus 0.4em\relax SPIE Press, 2010.

\bibitem{ref35}
M.~Nazarathy and J.~Shamir, ``Fourier optics described by operator algebra,''
  \emph{J. Opt. Soc. Am.}, vol.~70, no.~2, pp. 150--159, Feb 1980.

\bibitem{ref38}
P.~Yue, J.~Hu, X.~Yi, X.~Luan, and D.~Xu, ``Wave optics simulation
  investigation of multiple-input and aperture-averaging for optical wave
  propagation in turbulent ocean,'' \emph{Optics Communications}, vol. 452, pp.
  327--333, 2019.

\bibitem{ref39}
J.-R. Yao, M.~Elamassie, and O.~Korotkova, ``Spatial power spectrum of natural
  water turbulence with any average temperature, salinity concentration and
  light wavelength,'' \emph{Journal of the Optical Society of America A},
  vol.~37, no.~10, pp. 1614--1621, 2020.

\bibitem{ref41}
E.~Mohammed, U.~Murat, B.~Yahya, A.~Mohamed, and Q.~Khalid, ``Effect of eddy
  diffusivity ratio on underwater optical scintillation index,'' \emph{Journal
  of the Optical Society of America A Optics Image Science \& Vision}, vol.~34,
  no.~11, p. 1969, 2017.

\bibitem{ref42}
J.-R. Yao, H.-J. Zhang, R.-N. Wang, J.-D. Cai, Y.~Zhang, and O.~Korotkova,
  ``Wide-range prandtl/schmidt number power spectrum of optical turbulence and
  its application to oceanic light propagation,'' \emph{Opt. Express}, vol.~27,
  no.~20, pp. 27\,807--27\,819, Sep 2019.

\bibitem{ref34}
\BIBentryALTinterwordspacing
Teos-10: Thermodynamic equation of seawater-2010. [Online]. Available:
  \url{https://www.io-warnemuende.de/teos-10-2284.html}
\BIBentrySTDinterwordspacing

\bibitem{ref54}
T.~J. Mcdougall and P.~M. Barker, \emph{Getting started with TEOS-10 and the
  Gibbs Seawater (GSW) Oceanographic Toolbox}.\hskip 1em plus 0.5em minus
  0.4em\relax SCOR/IAPSO Working Group 127, 2011.

\bibitem{ref50}
A.~Goldman and J.~L. Devore, ``Probability and statistics for engineering and
  the sciences,'' \emph{Technometrics}, vol.~30, no.~2, p. 235, 1988.

\bibitem{ref51}
J.~H. Churnside, ``Aperture averaging of optical scintillations in the
  turbulent atmosphere,'' \emph{Applied Optics}, vol.~30, no.~15, pp.
  1982--1994, 1991.

\bibitem{ref52}
Z.~Wang, Q.~Wang, W.~Huang, and Z.~Xu, \emph{Visible Light Communications:
  Modulation and Signal Processing}.\hskip 1em plus 0.5em minus 0.4em\relax
  Wiley, 2017.

\bibitem{ref49}
A.~P. Prudnikov, Y.~A. Brychkov, and O.~I. Marichev, \emph{Integrals and
  Series, Special Functions}.\hskip 1em plus 0.5em minus 0.4em\relax Gordon and
  Breach Science Publishers S. A., London, UK, 1992.

\end{thebibliography}

\end{document}